\documentclass[epsfig,psfig,aps,twocolumn,prb,showpacs]{revtex4-1}
\usepackage{amsfonts,amsmath}
\usepackage{amsfonts}
\usepackage{graphicx}
\usepackage{epsfig}
\usepackage{relsize}
 
\begin{document}
%version without the supplemental material
\title
{
%Twisted bilayer graphene: interplay between twist and strain \\ 
Twistronics versus straintronics in twisted bilayers of graphene and transition metal dichalcogenides
}
\author
{
Marwa Manna\"{i} and Sonia Haddad$^{\ast}$
} 
\affiliation{
Laboratoire de Physique de la Mati\`ere Condens\'ee, D\'epartement de Physique,
Facult\'e des Sciences de Tunis, Universit\'e Tunis El Manar, Campus Universitaire 1060 Tunis, Tunisia
}
%\maketitle
%\date{\today}
%
%---------Abstract---------
%

\begin{abstract}
Several numerical studies have shown that the electronic properties of twisted bilayers of graphene (TBLG) and transition metal dichalcogenides (TMDs) are tunable by strain engineering of the stacking layers. In particular, the flatness of the low-energy moiré bands of the rigid and the relaxed TBLG was found to be, substantially, sensitive to the strain. However, to the best of our knowledge, there are no full analytical calculations of the
effect of strain on such bands. We derive, based on the continuum model of moiré flat bands, the low-energy Hamiltonian of twisted homobilayers of graphene and TMDs under strain at small twist angles. We obtain the analytical expressions of the strain-renormalized Dirac velocities and explain the role of strain in the emergence of the flat bands. We discuss how strain could correct the twist angles and bring them closer to the magic angle $\theta_m\sim1.05^{\circ}$ of TBLG and how it may reduce the widths of the lowest-energy bands at charge neutrality of the twisted homobilayer of TMDs. The analytical results are compared with numerical and experimental findings and also with our numerical calculations based on the continuum model.
\end{abstract}

%\keywords{...}
\maketitle

{\it Introduction.} Twistronics has, recently, emerged as a powerful tool to tailor the electronic properties of two-dimensional (2D) moiré systems, consisting of two accurately stacked layers of 2D materials with a relative twist angle $\theta$ \cite{Mc11,MacDo-Rev,Castro,Rev,Neto,Koshino15,Weck,Tutuc,Koshino18,Toma,March,Vish,Balents,Kaxiras,Macdo18,Macdo19,Naik18,Naik19,Pan18}. The first engineered two layer van der Waals structure is twisted bilayer graphene (TBLG), which has stimulated extensive theoretical and experimental studies, since the discovery of its superconducting state at $1.7 \rm{K}$ around the so-called magic angle (MA)$\theta_m\sim1.05^{\circ}$ \cite{Herrero1,Herrero2,Yank}.
This discovery has revived hope in unveiling the mechanism of superconductivity in high-$T_c$ (HTC) materials, which is one of the long standing puzzles of strongly correlated electrons systems.
The origin of superconductivity in TBLG is still under debate, but there is a general consensus on its extreme sensitivity to the occurrence, at the MA, of flat electronic bands, with $\mathrm{meV}$ width, located around the charge neutrality point and characterized by a vanishing effective Fermi velocity \cite{Volovik,Senthil,Wu,Roy,Bernevig,Efetov,Young,Herrero3}.
Flat bands have been, also, predicted to emerge in twisted bilayer transition metal dichalcogenides t-BTMD \cite{Naik18} and have been, recently, observed, over a wide range of twist angles \cite{Roy2}.
In TBLG, flat bands are found to emerge under a small uniaxial heterostrain, of $0.35\%$ at a relative twist angle $\theta\sim 1.25^{\circ}$ \cite{Bi,Guy}, which opens the way for strain engineering of flat bands \cite{Bi}.
Strain has also been found to be useful to probe the symmetry of the superconducting order parameter in TBLG \cite{SC-strain}.
Several experimental and numerical studies have, then, recently focused on the effect of strain on the electronic properties of TBLG and bilayers of transition metal dichalcogenides (TMDs) \cite{He13,Hung,Qiao,shear,SC-strain,Shi,strain-value,Paco,Hall,strainfield,He2,Guy2,Muller,Johnson,sara}.

However, a full analytical analysis of the strain dependence of the moiré band structure is still lacking.\\

In this Letter, we derive a low-energy effective Hamiltonian of TBLG and t-BTMD subject to an heterostrain. We determine the strain dependence of the effective Fermi velocities of the moiré flat bands. We also discuss the interplay between lattice relaxation and strain in TBLG.
The main result of the present work is that strain could be tuned in TBLG and t-BTMD to reduce the widths of the lowest-energy bands around charge neutrality. This finding may provide a platform to stabilize strongly correlated phases over a wider range of twist angles above the MA of TBLG and the critical angles of t-BTMD below which the bands could be regarded as flat.
Our results shed light on the experimental findings reporting a strain-induced MA in TBLG at small angle ($\theta\sim 1.25^{\circ}$) under a moderate heterostrain ($\epsilon\sim 0.3\%$) \cite{Qiao,Guy}.
Furthermore, we found that strain could counterbalance the effect of the lattice relaxation. The AA-stacking domains of the moiré structure are, then, expected to widen under strain, which furthers the emergence of the superconducting state, since the local density of states (LDOS) is peaked in these domains.
This is consistent with a recent experimental study reporting evidence of strain-induced strongly correlated phases in TBLG \cite{He2}.

Concerning the t-BTMDs, we found that the absence of MA in t-BTMDs is tied to their intralyer potential and interlayer tunneling parameters, which may be used to tailor a TMD van der Waals heterostructure with MA flat bands. 
Our results provide a tool to measure the strain tensor components of moiré systems, based on an accurate rotation of the layers.
The present work may, then, pave the way to a tunable strain moiré flat bands in 2D homobilayer materials.\\

{\it Continuum model of strained twisted bilayer graphene.} 
The low-energy Dirac Hamiltonian of monolayer graphene(MLG), rotated by an angle $\theta$ with respect to a fixed coordinate system and subject to a uniform strain, can be written as \cite{Bi,Oliva}:
\begin{eqnarray}
 h(\mathbf{k})=-\hbar v_F \left(\mathbb{I}+\mathcal{E}^T_t-\beta \mathcal{E}\right) \left(\mathbf{k}-\mathbf{D}_{\xi} \right)
 \cdot\mathbf{\sigma}^{\ast}
 \label{hML}
\end{eqnarray}
where $\xi$ is the valley index, $\mathbf{\sigma}^{\ast}=\left(\xi \sigma_x,\sigma_y\right)$ are the Pauli matrices, $v_F$ is the Fermi velocity of the undeformed layer and $\mathcal{E}_t=\mathcal{E}+R(\theta)$ is the total deformation tensor including the strain tensor $\mathcal{E}$ and the small-angle rotation matrix $R(\theta)$ written as:
\begin{eqnarray}
\mathcal{E}= 
\begin{pmatrix}
\epsilon_{xx} & \epsilon_{xy}\\
\epsilon_{xy} & \epsilon_{yy} 
\end{pmatrix},  \quad
R(\theta)= 
\begin{pmatrix}
0 & -\theta \\
\theta & 0 
\end{pmatrix}.
\end{eqnarray}
$\mathbf{D}_{\xi}=\left(\mathbb{I}-\mathcal{E}^T_t\right)\mathbf{K}^0_{\xi}-\xi \mathbf{A}$ is the position of the Dirac points \cite{Bi}, $\mathbf{K}^0_{\xi}=-\xi \frac{4\pi}{3a}\left(1,0\right)$ being the Dirac point of the undeformed layer, $\mathbf{A}=\frac{\sqrt{3}}{2a}\beta \left(\epsilon_{xx}-\epsilon_{yy},-2\epsilon_{xy}\right)$ is the effective gauge field, $\beta\sim 3$ for graphene \cite{Koshino17} and $a$ is the lattice parameter. 
The $\beta$ term in Eq. [\ref{hML}], which was not included in previous works \cite{Bi}, is due to the strain dependence of the inplane hopping parameters \cite{Oliva}. Actually, our numerical calculations show that this term could be neglected in TBLG regarding the small values of the strain amplitudes reported experimentally and which vary, for uniaxial deformation, from $0.1$ to $0.7\%$ \cite{strain-value}.\

We consider, as in Ref. \onlinecite{Bi}, a homobilayer AA stacking, where layers $1$ and $2$ are rotated and strained oppositely to preserve the orientation of the moiré Brillouin zone \cite{Jung}, with total deformation matrices $\mathcal{E}_{t2}=-\mathcal{E}_{t1}=\frac 12 \mathcal{E}_t$ where $\mathcal{E}_t=\mathcal{E}_{t2}-\mathcal{E}_{t1}$ is the relative deformation. \newline
We consider heterostrain since homostrain, where both layers are subject to identical strain, is found to slightly affect the moiré structure \cite{Guy}.\

We focus on the AA-stacked domains showing the largest LDOS and the highest electric conductivity compared with the AB/BA-stacked regions \cite{Zhang,Gadelha}. First, we study the rigid TBLG, and then we include the lattice relaxation effects.\

\begin{figure}[hpbt] 
\begin{center}
$\begin{array}{cc}
\includegraphics[width=0.45\columnwidth]{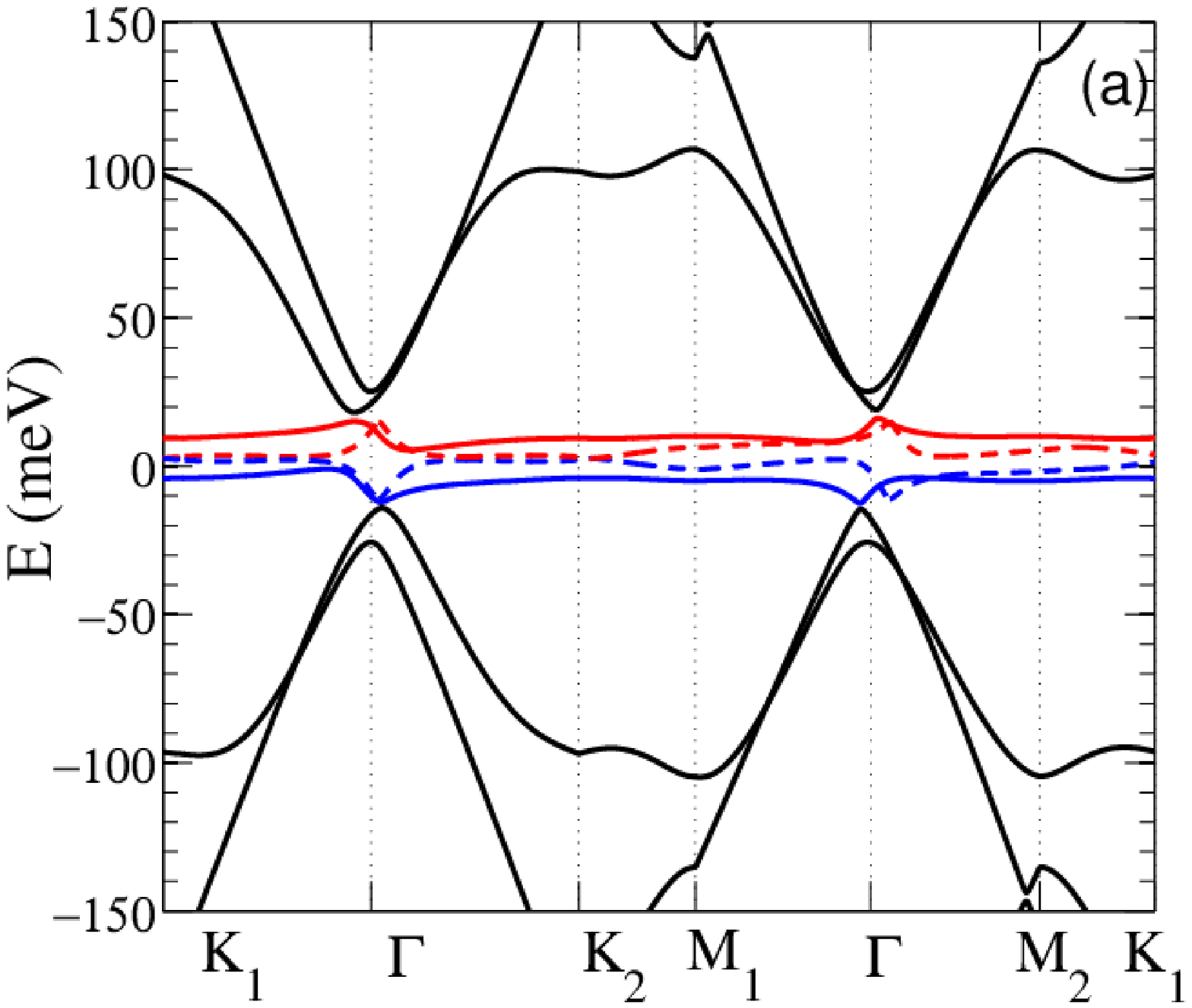}&
\includegraphics[width=0.45\columnwidth]{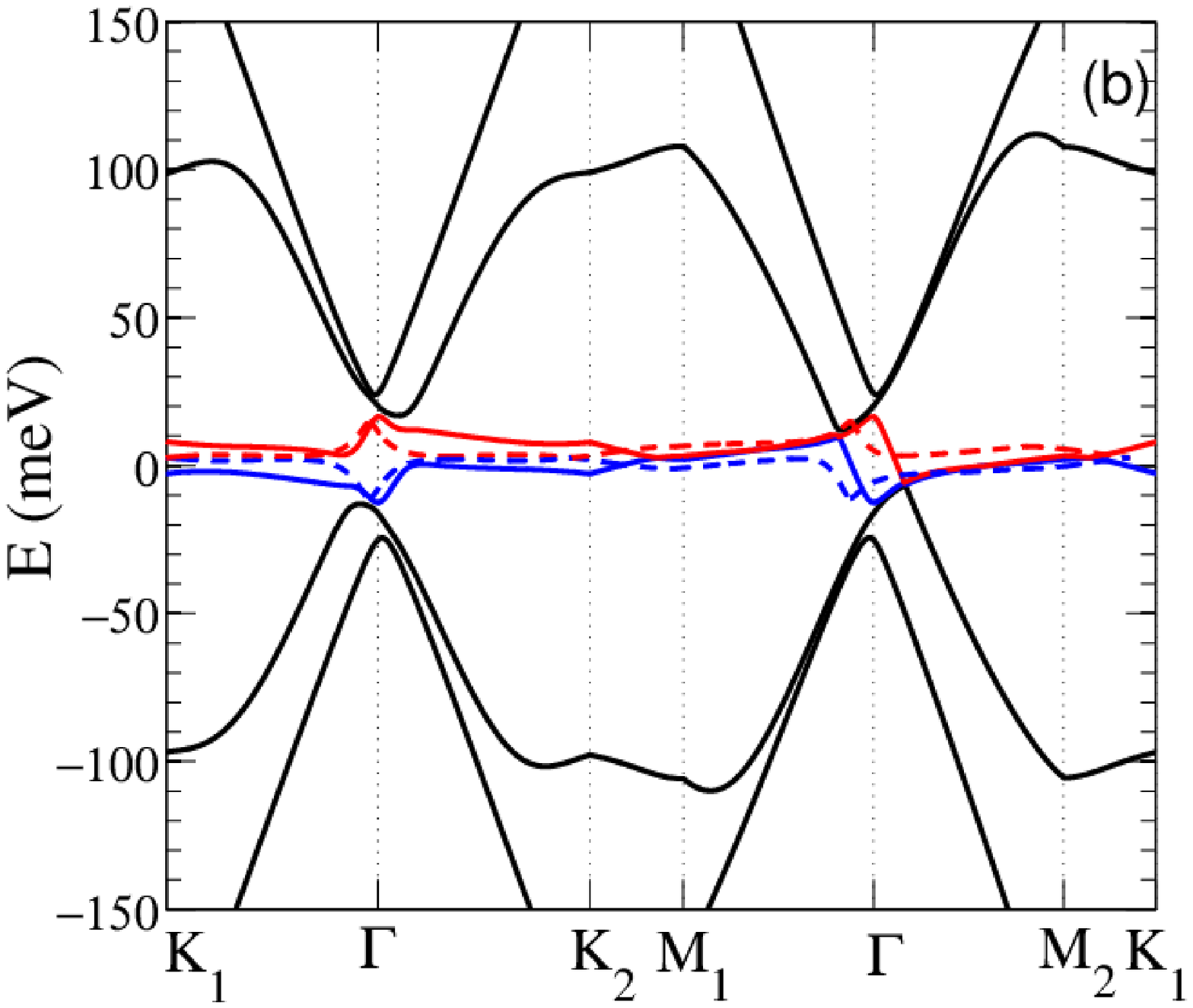}
\end{array}$
$\begin{array}{cc}
\includegraphics[width=0.45\columnwidth]{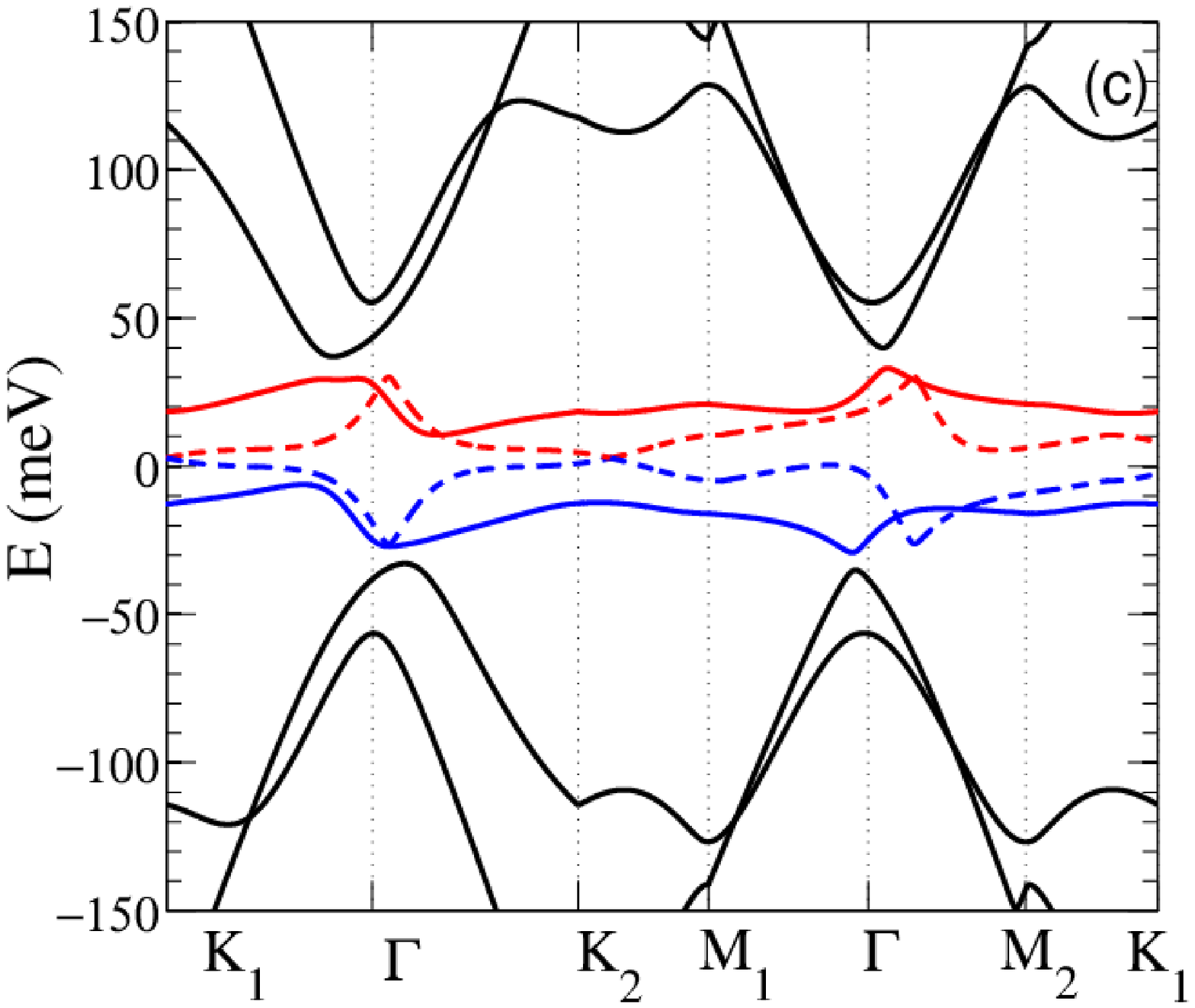}&
\includegraphics[width=0.45\columnwidth]{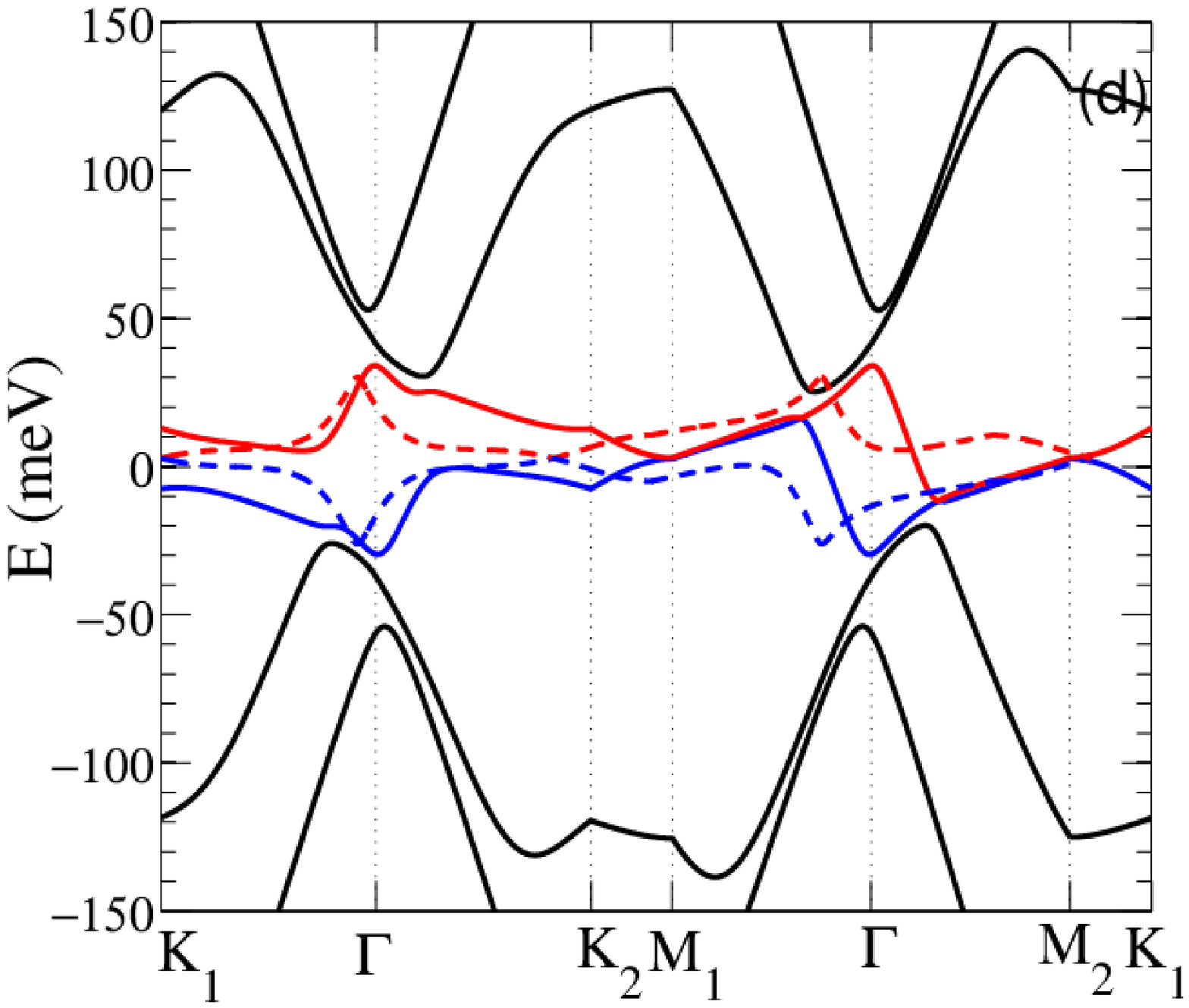}
\end{array}$
$\begin{array}{ccc}
\includegraphics[width=0.35\columnwidth]{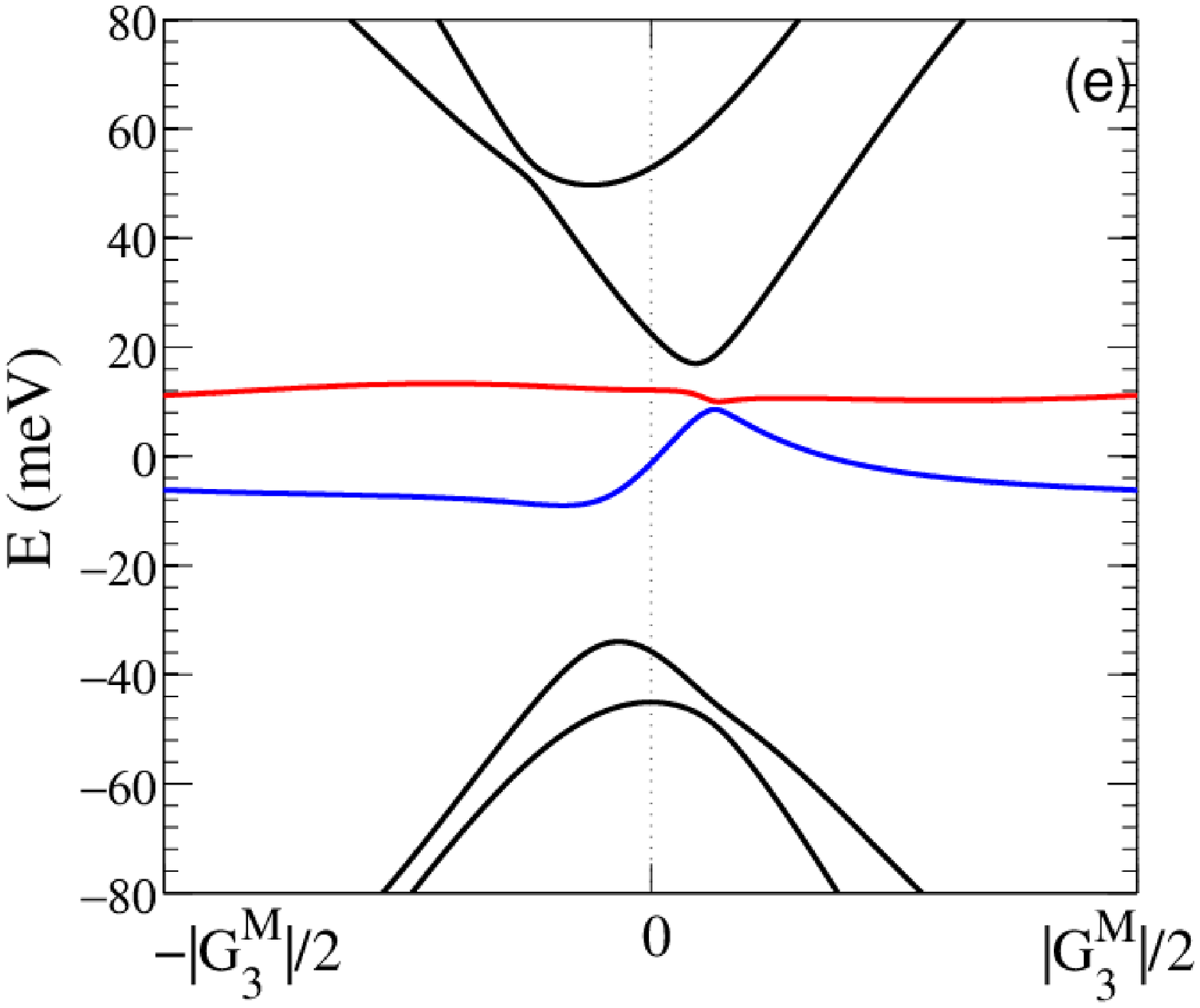}
\includegraphics[width=0.3\columnwidth]{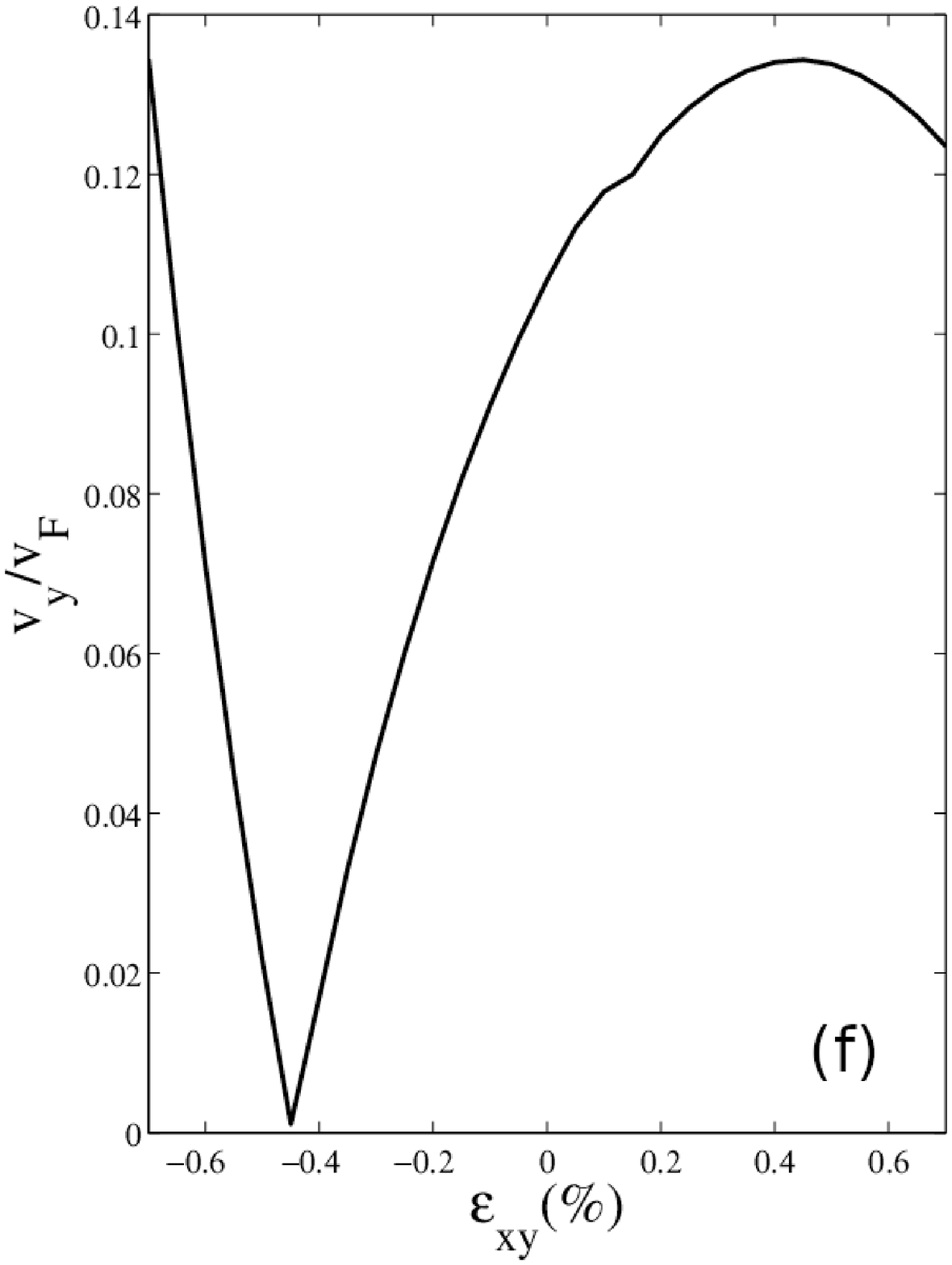}
\includegraphics[width=0.3\columnwidth]{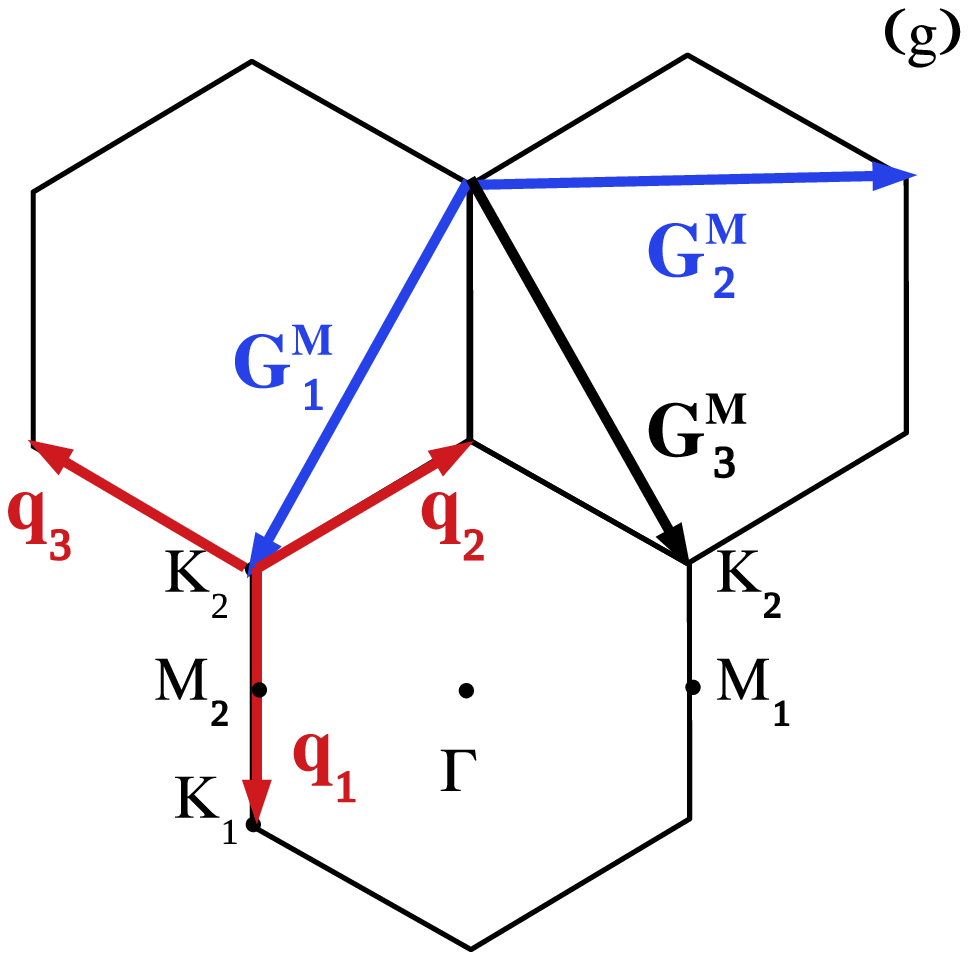}
\end{array}$
\end{center}
\caption{(a) Low-energy band structure of TBLG at twist angles $\theta=1.15^{\circ}$ [(a) and (b) and $\theta=1.25^{\circ}$ [(c) and (d)] without strain (dashed lines) and under a shear strain (solid lines) of $-0.16\%$  (a), $0.16\%$ (b), $-0.36\%$ (c) and $0.36\%$ (d). The bands are depicted in the deformed moir\'e Brillouin zone along the directions represented, in the unstrained case, in (g). Calculations are done for $w=118$ meV and $\hbar v_F/a\sim 2.68$ eV which corresponds to taking $\theta_m=1.05^{\circ}$ for the first MA. Here $a$ is the graphene lattice parameter. The band structures are calculated by diagonalizing the Hamiltonian given by Eq. [\ref{HBL}] in the basis of $\left\{|\mathbf {k}\rangle_1,|\mathbf{k+q_j}\rangle_2\right\}$ constructed by the states around, the Dirac point $\mathbf{D}_{1}$ of layer $1$ and those in the vicinity of the Dirac point $\mathbf{D}_{2}$ of layer $2$, respectively. A minimum number of 128 states is required to achieve the convergence for the low-energy bands. (e) Low energy bands along $\mathbf{G}^{M}_3=\mathbf{G}^{M}_1+\mathbf{G}^{M}_2$ and across the $\mathbf{D}_{1,\xi=-}$ point, where the moir\'e bands cross. This band structure is obtained for $\theta=1.15^{\circ}$ under a tensile shear strain $\epsilon_{xy}=0.16\%$ and for $w=118$ meV. (f) Strain dependence of the absolute value of the effective velocity $v_y$ [Eq.\ref{v*}] at a point along the $K_2M_1$ direction showing the largest velocity amplitude under a twist angle of $\theta=1.15^{\circ}$. $v_y$ vanishes at a compressive strain $\epsilon_{xy}\sim -0.4\%$, in agreement with the analytical value of $\epsilon_{xy}\sim -0.36\%$ ( Eq. [\ref{correction}]).}
\label{bandshear}
\end{figure}

Following the approach of Bistritzer and MacDonald \cite{Mc11} in deriving the low-energy continuum model of unstrained TBLG, we write the Hamiltonian of the strained TBLG as \cite{supp}
\begin{eqnarray}
H(\mathbf{k})= 
\begin{pmatrix}
h_1(\mathbf{k}) & T_1 & T_2 & T_3\\
T^{\dagger}_1 & h_{2,1}(\mathbf{k}) & 0&0\\
T^{\dagger}_2 & 0& h_{2,2}(\mathbf{k}) & 0\\
T^{\dagger}_3 & 0& 0& h_{2,3}(\mathbf{k})\\
\end{pmatrix},
\label{HBL}
\end{eqnarray}
where the corresponding basis $\Psi=\left(\psi_0(\mathbf{k}),\psi_1(\mathbf{k}),\psi_2(\mathbf{k}),\psi_3(\mathbf{k})\right)$ is constructed on the two-component sublattice spinor $\psi_0(\mathbf{k})$ ($\psi_j(\mathbf{k})$) of layer $1$ (layer $2$) taken at the momentum $\mathbf{k}$ measured from $\mathbf{D}_{1\xi}$  at a given valley $\xi$. 
$h_1\left(\mathbf{k}\right)=-\hbar v_F  \left(\mathbb{I}+ \mathcal{E}_{t1}-\beta\mathcal{E}_1\right)\mathbf{k}\cdot\mathbf{\sigma}^{\ast}$ is the Hamiltonian of layer $1$ written in the vicinity of $\mathbf{D}_{1\xi}$ and $h_2\left(\mathbf{k}\right)=-\hbar v_F \left(\mathbb{I}+ \mathcal{E}_{t2}-\beta\mathcal{E}_2\right) \left(\mathbf{k}+\mathbf{q}_{j\xi}\right)\cdot\mathbf{\sigma}^{\ast}$, $(j=1,2,3)$ is that of layer $2$ written around $\mathbf{D}_{2\xi,j}$ where $\mathcal{E}_{ti}=\mathcal{E}_i+R(\theta_i)$ is the total deformation tensor of layer $i$. The $\mathbf{q}_{j\xi}$ vectors connecting the Dirac points $\mathbf{D}_{2\xi,j}$ to $\mathbf{D}_{1\xi}$ are 
$\mathbf{q}_{1\xi}=\mathbf{D}_{1\xi}-\mathbf{D}_{2\xi}$, 
 $\mathbf{q}_{2\xi}=\mathbf{q}_{1\xi}+\xi\mathbf{G}^M_1$, and $\mathbf{q}_{3\xi}=\mathbf{q}_{1\xi}+\xi\left(\mathbf{G}^M_1+\mathbf{G}^M_2\right)$,
where, $\left(\mathbf{G}^M_1,\mathbf{G}^M_2\right)$ is the moiré Brillouin zone (BZ) basis given by 
$\mathbf{G}^M_i=\mathcal{E}_t^T\mathbf{G}_i$ 
where $\mathbf{G}_1=\frac{2\pi}a\left(1,-1/{\sqrt{3}}\right)$ and$\mathbf{G}_2=\frac{2\pi}a\left(0,2/{\sqrt{3}}\right)$ are the reciprocal lattice vectors associated with the undeformed monolayer lattice basis constructed on the primitive lattice vectors $\mathbf{a}_1=a\left(1,0\right)$ and $\mathbf{a}_2=a\left(1/2,\sqrt{3}/2\right)$ (Fig. \ref{bandshear}).\newline
In the absence of strain, the $\mathbf{q}_{j\xi}$ vectors satisfy $\sum_j \mathbf{q}^0_{j\xi}=\mathbf{0}$, where $\mathbf{q}^0_{j\xi}$ denote the corresponding vectors of the unstrained system. We set hereafter 
\begin{eqnarray}
 \mathbf{q}_{j\xi}=\mathbf{q}^0_{j\xi}+\Delta\mathbf{q}_{j\xi},
 \label{qj}
\end{eqnarray}
where $\Delta\mathbf{q}_j$ is the strain-induced correction\cite{supp}.\

For an AA bilayer stacking, the $T_j$ matrices reduce to $T_1= w\left(\mathbb{I}+\sigma_x\right)$, 
$T_2=w \left(\mathbb{I}-\frac 12\sigma_x+\xi \frac {\sqrt{3}}2\sigma_y\right)$, and $T_3=w\left( \mathbb{I}-\frac 12\sigma_x-\xi\frac {\sqrt{3}}2\sigma_y\right)$ \cite{supp}, where we assumed, for simplicity, that the interlayer tunneling amplitude $w\sim 118\, \mathrm{meV}$ is strain independent. This assumption is justified since the vertical interlayer hopping parameter is found to be, to the first order in strain, unchanged compared with the undeformed lattice \cite{Falko20}.
We also consider rigid TBLG where the interlayer tunneling amplitudes in the AA, AB/BA, and BB regions are assumed to be equal. The interplay between strain and lattice relaxation will be discussed later.\

Regarding the small values of the twist angle $\theta$ and the strain amplitudes in TBLG \cite{strain-value}, a low-energy Hamiltonian $H^{(1)}(\mathbf{k})$ can be derived from Eq. [\ref{HBL}], based on a first-order perturbative approach as done in Ref. \onlinecite{Mc11}. \
To the leading order in $\mathbf{k}$, $H^{(1)}\left(\mathbf{k}\right)$ can be written as

\begin{widetext}
\begin{eqnarray}
 H^{(1)}\left(\mathbf{k}\right)=\frac{\langle\Psi|H(\mathbf{k})|\Psi\rangle}{\langle\Psi|\Psi\rangle}=\frac1{\langle\Psi|\Psi\rangle}
 \left[\psi^{\dagger}_0h_0\left(\mathbf{k}\right) \psi_0+ \psi^{\dagger}_0\sum_jT_j h^{-1}_j h_0\left(\mathbf{k}\right)h^{-1}_j T^{\dagger}_j \psi_0\right]
 \label{H1}
\end{eqnarray}
\end{widetext}
$\psi_0$ being a zero-energy state of the monolayer Hamiltonian $h_0$ (Eq. [\ref{hML}]) where the strain and the twist angle could be neglected \cite{Mc11,supp}. $h_j=h_{2,j}(\mathbf{k}=\mathbf{0})\sim -\hbar v_F \mathbf{\sigma}^{\ast}\cdot\mathbf{q}_{j\xi}$, where $\mathbf{q}_{j\xi}$ are given by Eq. [\ref{qj}] \cite{supp}.\\

To the leading order in strain amplitude, the four two-component spinor $\Psi$ satisfies, as in the unstrained case, $\langle\Psi|\Psi\rangle\sim1+6\alpha^2$, where $\alpha=\frac{w}{\hbar v_F k_{\theta} }$ and $k_{\theta}=2K^0\sin \theta/2\sim \frac{4\pi}{3a}\theta$, $K^0=\frac{4\pi}{3a}$ is the amplitude of the Dirac point vector of the undeformed layer\cite{supp}.
The low-energy $2\times 2$ Hamiltonian of Eq. [\ref{H1}] reduces, then, to
\begin{widetext}
 \begin{eqnarray}
 H^{(1)}\left(\mathbf{k}\right)=-\frac{\hbar}{1+6\alpha^2}\psi^{\dagger}_0\left[ v_{0x} k_x+v_{0y} k_y+ \xi \sigma_x v_x k_x +
 \sigma_y v_y k_y+\xi \sigma_x v_{xy}k_y+\sigma_y v_{yx}k_x\right] \psi_0
 \label{H1strain}
\end{eqnarray}
\end{widetext}
where the tilt parameters $\mathbf{v_0}=(v_{0x},v_{0y})$, and the strain-renormalized velocities are given by
\begin{widetext}
 \begin{eqnarray}
 &&v_{0x}=-\xi v_F \frac{16\pi}a\frac{\alpha^2}{k_{\theta}}\epsilon_{xy},\; v_{0y}=-\xi v_F \frac{8\pi}a\frac{\alpha^2}{k_{\theta}}\left(\epsilon_{xx}-\epsilon_{yy}\right), \nonumber\\
 &&v_{x}=v_F\left(1-3\alpha^2-\frac {6\sqrt{3}}a\frac{\alpha^2}{k_{\theta}}\beta \epsilon_{xy}\right) \;
 v_{y}=v_F\left(1-3\alpha^2+\frac {6\sqrt{3}}a\frac{\alpha^2}{k_{\theta}}\beta \epsilon_{xy}\right),\nonumber\\
 && v_{xy}= v_F \frac{\alpha^2}{k_{\theta}}\left[\left(\frac{3\sqrt{3}}a\beta-\frac{4\pi}{a}\right)\epsilon_{xx}
 -\left(\frac{4\pi}{a}+\frac{3\sqrt{3}}a\beta\right)\epsilon_{yy}\right],\;
 v_{yx}= v_F \frac{\alpha^2}{k_{\theta}}\left[\left(\frac{4\pi}{a}+\frac{3\sqrt{3}}a\beta\right)\epsilon_{xx}
 + \left(\frac{4\pi}{a}-\frac{3\sqrt{3}}a\beta\right)\epsilon_{yy}\right]\nonumber\\
 \label{v*}
\end{eqnarray}
\end{widetext}

Equations [\ref {H1strain}] and [\ref{v*}], which are one of the main results of the present work, reduce to the expressions obtained by Bistritzer and MacDonald \cite{Mc11}, in the limit of a vanishing strain.\

In the following, we discuss the effect of strain on the flat bands, appearing around the MA.\\

{\it Flat-band behavior under strain.}
According to Eqs. [\ref{H1strain}] and [\ref{v*}], the flatness of the low-energy bands can be  selectively tuned by the strain along the moiré BZ directions by choosing, for a given twist angle $\theta$, the strain value at which the corresponding effective velocity vanishes.\

Considering a shear strain $\epsilon_{ij_\ne i}=\epsilon, \epsilon_{ii}=0$ \cite{shearML,shearML2}, the $v_{0y}$ tilt component and the cross velocity terms $v_{xy}$ and $v_{yx}$ turn to zero, while the velocities $v_x$ and $v_y$ along the $k_x$ and $k_y$ axes, respectively, read as
\begin{eqnarray}
 v_{x,y}=v^{\ast}_0 +\Delta v_{x,y}
\end{eqnarray}
where $v^{\ast}_0=v_F\frac{1-3\alpha^2}{1+6\alpha^2}$ is the low-energy effective velocity of the unstrained TBLG \cite{Mc11} and the strain-induced corrections are
\begin{eqnarray}
 \Delta v_{x,y}=\mp \frac {v_F} {1+6\alpha^2} \frac{\alpha^2}{k_{\theta}} \frac {6\sqrt{3}}a\beta \epsilon_{xy}.
 \label{v-correct}
\end{eqnarray}
According to Eq. [\ref{v*}], under a compressive (tensile) shear strain 
$\epsilon_{xy}<0$ ($\epsilon_{xy}>0$), the renormalized velocity $v_{y}$ decreases (increases) compared with the unstrained value $v^{\ast}_0$ (Fig. \ref{bandshear}). 
At a twist angle $\theta=1.25^{\circ}$, $v_{y}$ vanishes for a compressive shear strain of $-0.36\%$ amplitude, which is consistent with our numerical results depicted in Fig. \ref{bandshear}(f) showing 
the strain dependence of the effective velocity $v_{y}$ along the $K_2M_1$ direction of the moiré BZ represented in Fig. \ref{bandshear}(g). 
A flat band can, then, emerge in TBLG under an accurately applied strain at $\theta>\theta_m$, which may give rise to a strain-induced superconductivity over a wider range of twist angles and not only at the low MA, which should be accurately tuned to stabilize the strongly corrected phases.
A detailed analysis of the strain dependence of the superconducting critical temperature based on the approaches used in Refs. \onlinecite{haddad,CNT} are needed.\newline
Moreover, the tilt term $v_{0x}$ deforms the Dirac cone and breaks the particle-hole symmetry.
These features are in agreement with the numerical results of Fig. \ref{bandshear} (e) showing a deformed Dirac cone at the crossing of the low-energy bands around the charge neutrality point. 
It is worth stressing that the effect of strain, on the moiré bands of TBLG, goes beyond the reduction in the flat band width. Heterostrain, near the magic angle, was found to generate a zero-energy flat band between the two van Hove singularities and a valley degeneracy lifting \cite{He2}.\

%\begin{widetext}
%\end{widetext}
%%%%%%%%%%%%%%
The strain-modified Dirac cone shapes could affect the electron-phonon interactions.
In monolayer graphene, the strain-induced tilt of Dirac cones has been found to, particularly, affect the Kohn anomaly \cite{haddad}. Such an anomaly has also been observed in BLG and may, also be, sensitive to strain and twist. This point needs to be studied further.\

The opposite signs of $\Delta v_{x,y}$, in Eq.[\ref{v-correct}], are due to the off-diagonal structure of the shear strain tensor, which is similar to the small-twist-angle $R(\theta)$ matrix but with the same sign for both strain tensor components. Therefore a twist gives rise to an isotropic velocity $v^{\ast}=v_F(1-3\alpha^2)/(1+6\alpha^2)$ \cite{Mc11}, while a shear strain leads to anisotropic velocities with opposite corrections. 
It is worth stressing that these corrections are only due to the gauge field $\mathbf{A}$. This means that, under a shear strain, the displacement of the Dirac points is the key parameter governing the flatness of the moiré bands. 

In Ref. \onlinecite{shear}, the authors reported, based on density functional theory (DFT) calculations, that a shear deformation at an angle $\gamma\sim 0.08^{\circ}$ along the armchair direction introduces a correction of $\Delta\theta_m=0.04^{\circ}$ to the calculated MA, $\theta^{th}_m=1.12^{\circ}$, to agree with the experimental value of $\theta_m\sim1.08^{\circ}$. This correction can be understood from the expressions of the renormalized velocities given by Eq. [\ref{v*}]. Let us denote as $\alpha^{\ast}$ the corrected value of $\alpha=\frac{w}{\hbar v_F k_{\theta}}$ defined by
\begin{eqnarray}
v_{x,y}=v_F\left(1-3\alpha^2\right)+\Delta v_{x,y}\equiv v_F\left(1-3\alpha^{\ast 2}\right),
\label{correction}
\end{eqnarray}
For a twist angle $\theta^{th}_m=1.12^{\circ}$ and a shear strain $\epsilon\sim \tan\gamma\sim \gamma=0.14\%$, and taking $\hbar v_F/a=2.68\, \mathrm{eV}$ for graphene \cite{Bi}, we deduce, from Eq.[\ref{correction}], that the corrected MA, ascribed to $\alpha^{\ast}$, is $\theta^{\ast}=1.04^{\circ}$, which is in good agreement with the measured MA of $1.05^{\circ}$.\

For a uniaxial strain along the armchair or zigzag directions, the velocity corrections $\Delta v_{x,y}$ vanishe, which may be used as a probe to check the direction of the applied strain.\

Let us consider, as in Ref. \onlinecite{Bi}, a twist angle of $\theta=1.05^{\circ}$ and a uniaxial strain of $\epsilon=0.6\%$ applied along the direction $\varphi=30^{\circ}$.
According to Eqs. [\ref{H1strain}] and [\ref{v*}], the corrections to the Dirac velocities reach $\pm 0.13 v_F$ \cite{supp}, which agrees with the numerical calculations of Ref. \onlinecite{Bi}, where an increase of $0.14v_F$ was reported. \\

Based on structural and spectroscopic measurements combined with tight-binding (TB) calculations, Huder{\it et al.} \cite{Guy} showed that flat bands emerge in TBLG under a small uniaxial heterostrain, of the order of $0.35\%$, and a twist angle of $\theta=1.25^{\circ}$. 
Taking $\theta_m=1.05^{\circ}$ as the first MA, $v_y$ [Eq.\ref{v*}] vanishes for a uniaxial strain having an off-diagonal component $\epsilon_{xy}=-0.36\%$, as shown in Fig. \ref{bandshear} (f), which is consistent with the results of Ref. \onlinecite{Guy}.\

The outcomes of our results show that the Hamiltonian given by Eq. [\ref{H1strain}] may provide insights into the behavior of the low-energy bands of TBLG under strain. \\

{\it TBLG: Interplay between strain and lattice relaxation.} 
Several studies focused, recently, on the inclusion of lattice relaxation and deformation effects in the continuum model of TBLG \cite{Guinea,Koshino20,Koshino18,Carr,Fang19}. 
However, to the best of our knowledge, there is no analytical expression of the effective Fermi velocity of the relaxed-TBLG, which is an indicator of the flatness of the low-energy bands.
Here, we derive the renormalized velocities under strain, taking into account the relaxation effects.

\begin{figure}[hpbt] 
\begin{center}
$\begin{array}{cc}
\includegraphics[width=0.5\columnwidth]{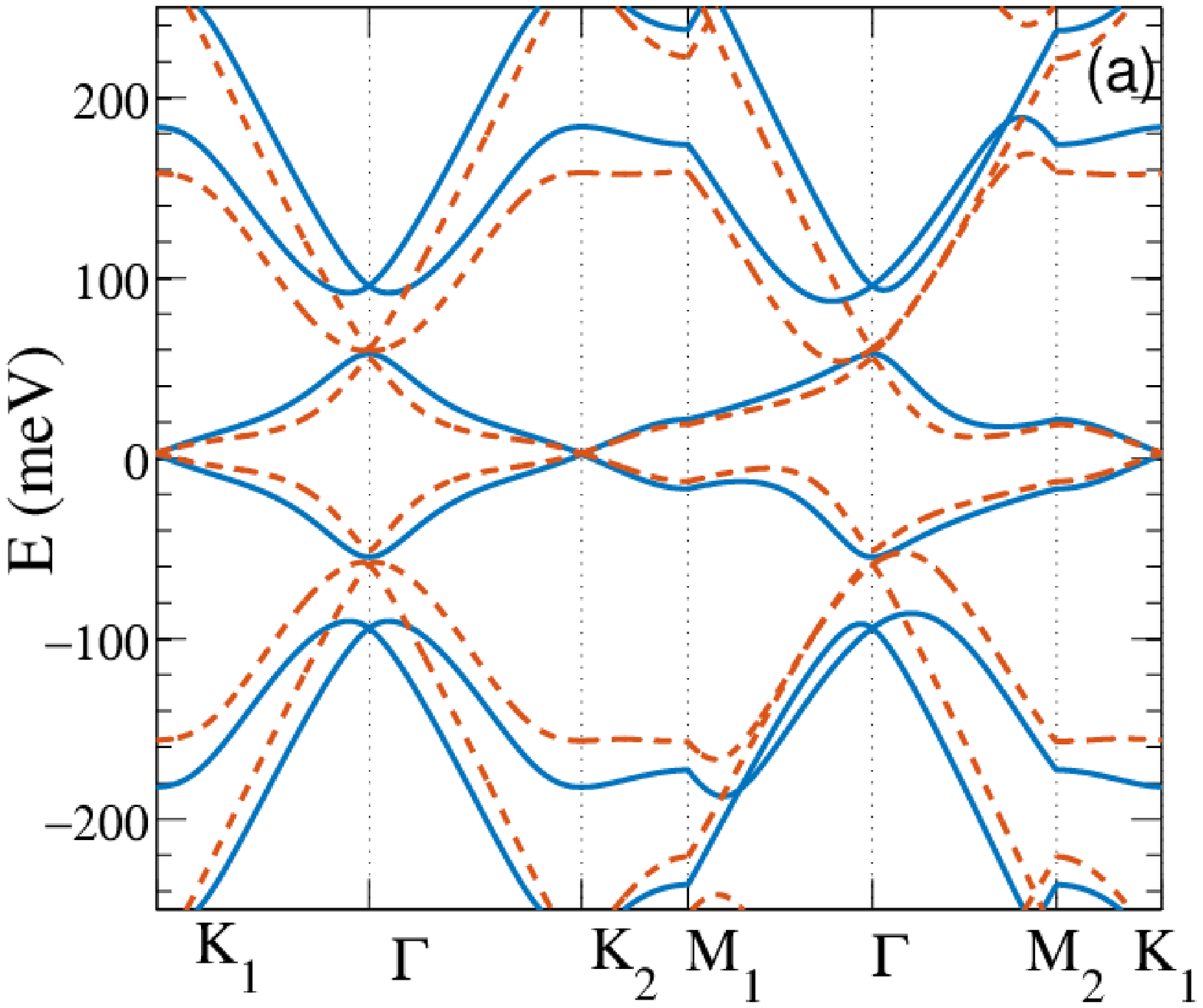}
\includegraphics[width=0.5\columnwidth]{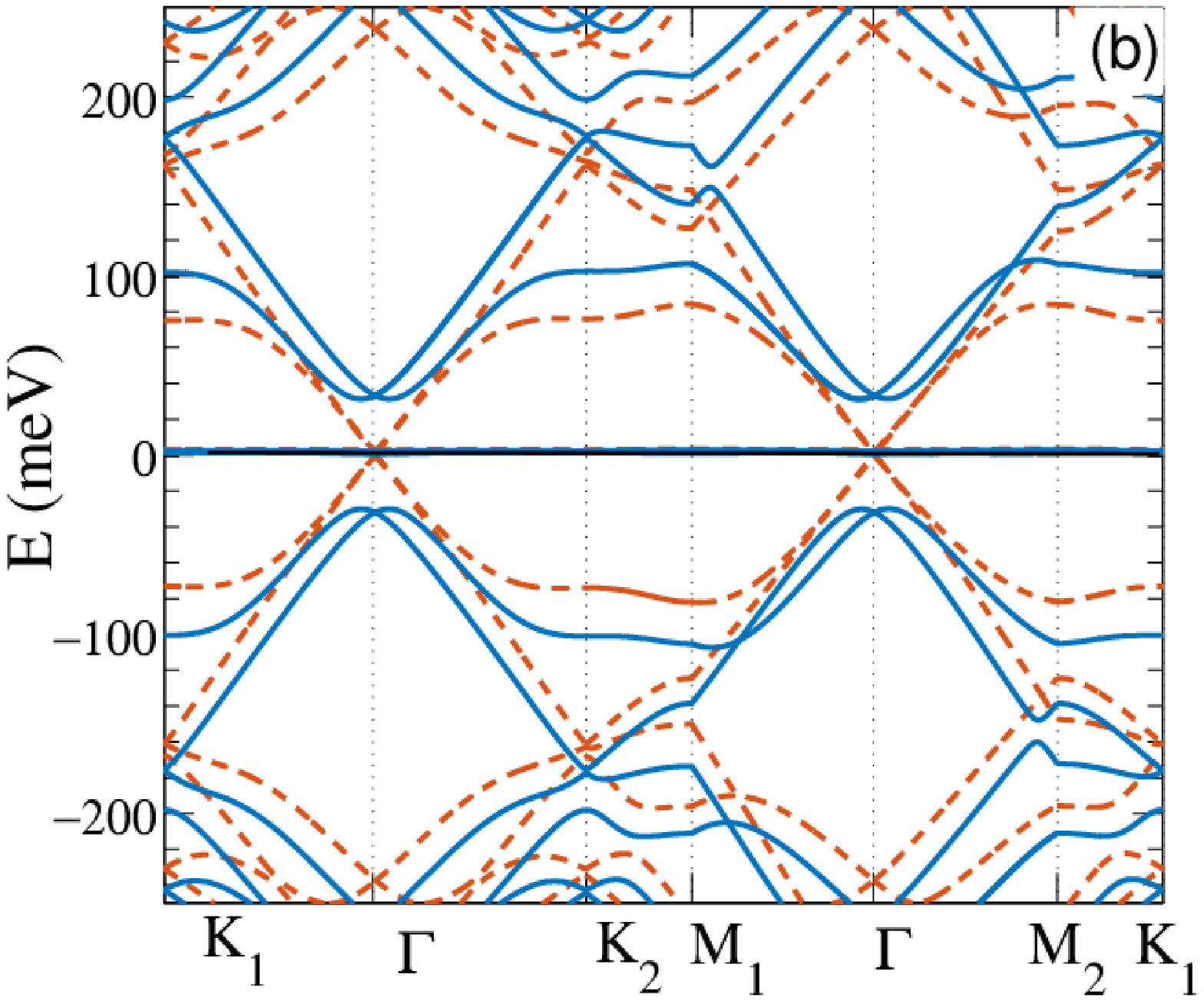}
\end{array}$
\includegraphics[width=0.5\columnwidth]{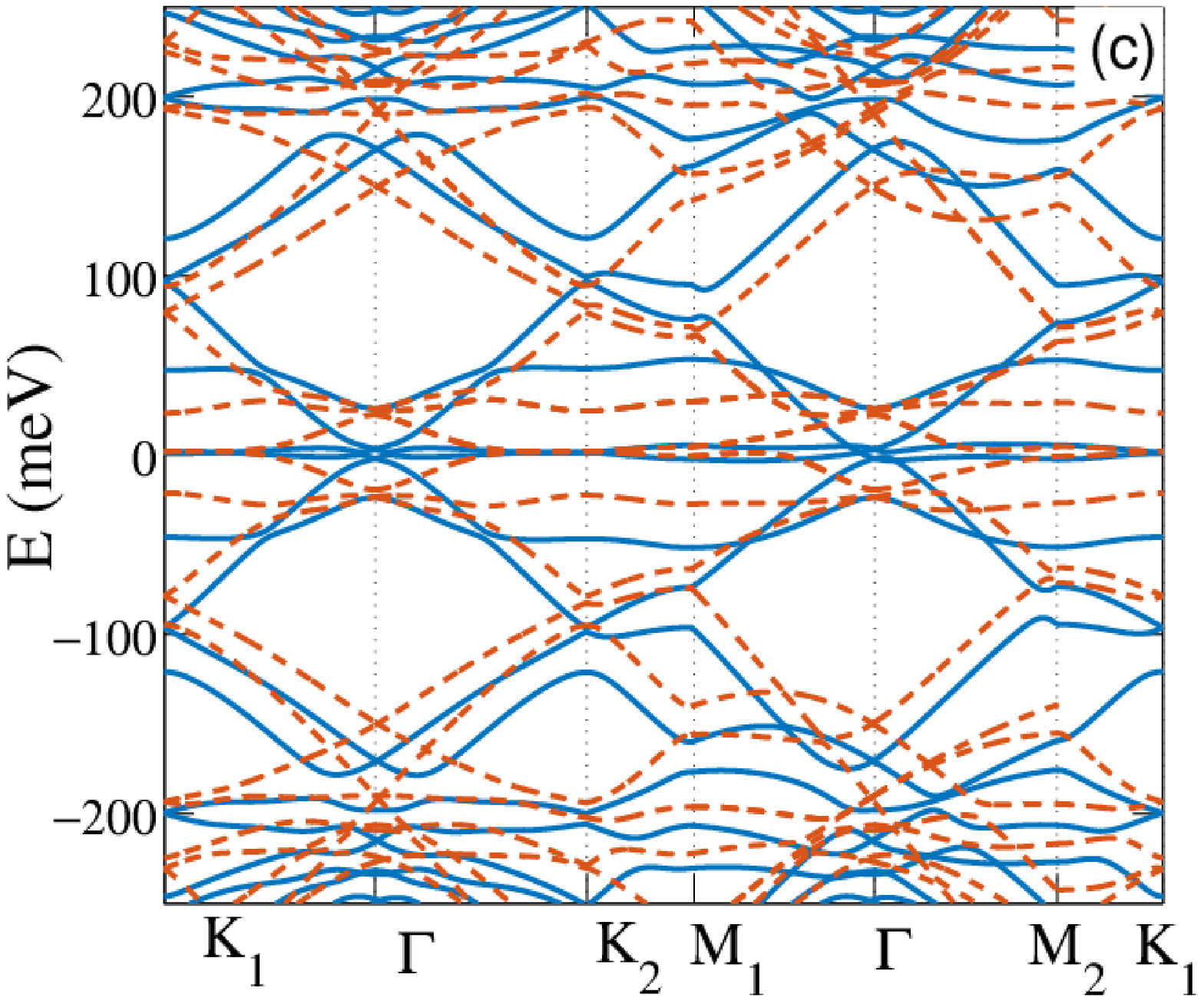}
\end{center}
\caption{ Low-energy band structure of unstrained TBLG at twist angle $\theta=1.4^{\circ}$ (a), $\theta=1.05^{\circ}$ (b)  and $\theta=0.8^{\circ}$ (c). The solid (dashed) lines correspond to relaxed (rigid) lattice. Calculations are done for $w=118$ meV in the unrelaxed case and for $w=w^{\prime\prime}94$ meV and for$w^{\prime}=118$ meV for the relaxed lattice \cite{Cantele}.}
\label{relax}
\end{figure}

The effect of relaxation can be included in the interlayer tunneling matrices $T_j$ by taking $w\sim 0.090$ eV and $w^{\prime}=0.117$ eV \cite{Sarma2,Koshino20}. 
For simplicity, we set $w^{\prime\prime}=w$ \cite{Jung} and consider AA-stacked regions. In this case, the $T_j$ matrices take the forms $T_1=w\left(\mathbb{I} +u \sigma_x\right), \, 
T_2= w\left[\mathbb{I} +u \left(-\frac 12\sigma_x+\xi\frac{\sqrt{3}}2\sigma_y\right)\right],$ and $T_3= w\left[\mathbb{I} +u \left(-\frac 12\sigma_x-\xi\frac{\sqrt{3}}2\sigma_y\right)\right]$,
where $u=\frac{w^{\prime}}{w}\sim1.25$ \cite{Koshino20,Cantele} is the ratio of the interlayer tunneling amplitudes, which we have assumed to be equal in the previous section dealing with rigid TBLG.\newline
Following the same procedure as the previous section, we derive the low-energy effective Hamiltonian of the relaxed TBLG under strain \cite{supp}. At low-energy, we found that the effective Fermi velocity of the relaxed TBLG in the absence of strain is
\begin{eqnarray}
v_r^{\ast}=v_F\frac{1-3u^2\alpha^2}{1+3\left(1+u^2\right)\alpha^2},
\label{vrelax}
\end{eqnarray}
According to Eq. [\ref{vrelax}], the MA in the relaxed lattice is larger than that in the unrelaxed case \cite{supp}. Moreover, the relaxation is found to reduce the bandwidth of the lowest bands at charge neutrality, which is consistent with the band structure, around MA, depicted in Fig. \ref{relax} and previous TB calculations \cite{Koshino20,Guinea,Cantele}. However, Eq. [\ref{vrelax}] does not account for the relatively large bands of the relaxed lattice at $\theta$ above the MA, as reported from first-principles and TB calculations \cite{Koshino20,Guinea,Cantele} and as shown in Fig. \ref{relax}, which limits the angle range of application of the $2\times 2$ low-energy Hamiltonian (Eq. [\ref{H1}]) in the case of relaxed TBLG. \

On the other hand, our results show that, at a given twist angle, a smaller strain amplitude is needed to flatten the lowest bands of the relaxed lattice compared with the rigid TBLG \cite{supp}.\

{\it Continuum model for t-BTMDs.} 
For a large family of TMDs, a strained monolayer can be described, to the lowest order of $\mathbf{k}$, by the gapped Dirac Hamiltonian given, in the basis of the atomic orbital states of the conduction, $|\Phi_c\rangle$, and valence, $|\Phi_v\rangle$, bands by \cite{Bi,Liu13,Xiao12,Fang18,Peeters}
\begin{eqnarray}
 h(\mathbf{k})&=&-\hbar v_F \left(\mathbb{I}+\mathcal{E}^T_t-\beta \mathcal{E}\right)\mathbf{k}
 \cdot \left(\xi \sigma_x,\sigma_y\right)\nonumber\\
 &+&\frac m2 \left(\sigma_z+\mathbb{I}\right),
 \label{h1TMD}
\end{eqnarray}
where $\mathbf{k}$ is measured with respect to the Dirac point $\mathbf{D}_{\xi}$. 

We consider a homobilayer TMD structure where the layers are rotated and strained oppositely with the same amplitude, as in the TBLG case. The corresponding low-energy Hamiltonian can be written in the form of Eq. [\ref{HBL}] by replacing the $h_i(\mathbf{k})$ by $h^{\prime}_i(\mathbf{k})=h_i(\mathbf{k})+V_i(\mathbf{r})$, $(i=1,2)$. $h_1$ ($h_2$) is the Hamiltonian of layer $1$ in the vicinity of the Dirac point $\mathbf{D}_{1\xi}$ ($\mathbf{D}_{2\xi}$). The Dirac points are connected by the three hopping processes $\mathbf{q}_{j\xi}$. $h_1$ is given by Eq. [\ref{h1TMD}] and $h_2=-\hbar v_F \left(\mathbb{I}+\mathcal{E}^T_{t2}-\beta \mathcal{E}_2\right) \left(\mathbf{k}+\mathbf{q}_{j\xi} \right)\cdot\left(\xi \sigma_x,\sigma_y\right)\nonumber\\
+\frac m2 \left(\sigma_z+\mathbb{I}\right)$. Moreover, one should include in Eq. (\ref{HBL}) a diagonal intralayer potential $V_l\left(\mathbf{r}\right)$, $(l=1,2)$ \cite{Bi,Macdo18,Jung}, which is the same for both layers in the AA-stacked t-BTMD \cite{supp}.\newline
We found that the effective Fermi velocity of the unstrained rigid lattice is given by \cite{supp}
\begin{eqnarray}
 v^{\ast}_{TMD}= \frac{v_F}{\langle\Psi|\Psi\rangle} \left(1+\frac{3w^2}{\gamma-\left(\hbar v_Fk_{\theta}\right)^2}\right),
 \label{v_TMD}
\end{eqnarray}
where $\gamma=U_v\left(m+U_c\right)$, $U_i=6V_i\cos\Phi_i, (i=v,c)$ are the inralayer potential parameters \cite{supp} and $\langle\Psi|\Psi\rangle$ is expressed in the supplemental material \cite{supp}.\

According to Eq. [\ref{v_TMD}], twisted TMD bilayers, with negative $\gamma$, as in WSe$_2$, do not show MA.
However, flat bands can emerge over a wide range of twist angles $\theta<\theta_c$, where $\theta_c$ is a critical value corresponding to a threshold bandwidth, below which the bands could be regarded as flat (Fig. 1 of the Supplemental Material \cite{supp}). This feature is in agreement with Ref. \onlinecite{Jung}.
It is worth stressing that Angeli and MacDonald \cite{Angeli} developed a continuum theory combined with {\it ab} initio calculations to account for the twist dependence of the moiré valence bands at the $\Gamma$ point of twisted homobilayer TMD. However, as far as we know, there are no analytical expressions of the twist dependence of the effective mass or the effective velocities of t-BTMD around the $\Gamma$ \cite{Angeli} or Dirac points \cite{Jung}.\

According to Eq. [\ref{v_TMD}], magic angles with vanishing effective Fermi velocity may occur in t-TMDs if $\gamma$ is positive, which provides a tool to tailor TMD materials with the MA by an accurate choice of the intralayer potential and the interlayer tunneling parameters.\\

The strain dependence of the effective velocity of t-BTMD \cite{supp} shows that the lowest bands could be flattened by strain for a twist angle above $\theta_c$.
Under a shear (uniaxial) deformation, the strain-induced correction to the $v_x$ velocity component is
$\Delta v^{\prime}_{x}\sim - 0.2\,\epsilon_{xy} v_F$ ($\Delta v^{\prime}_{x}=\sim 0.54 \,\epsilon_{xx} v_F$) at a twist angle $\theta=3^{\circ}$ \cite{supp}.
It comes out that, in contrast to TBLG, the flatness of the low-energy bands of t-BTMD is not considerably affected by the strain as depicted in  Fig. 2 of the Supplemental Material \cite{supp}, which represents the band structure of twisted homobilayer of WSe$_2$ calculated by diagonalizing the low-energy Hamiltonian of t-BTMD \cite{supp} in $128$-state basis of $\left\{|\mathbf {k}\rangle_1,|\mathbf{k+q_j}\rangle_2\right\}$.

%\end{widetext}
%%%%%
In conclusion, we derived a low-energy Hamiltonian of TBLG and t-BTMD under a strain deformation which captures the interplay between strain and twist and opens the way to control the flatness of the moiré bands. We determined the analytical expressions of the strain-induced corrections of the effective Fermi velocities. Our results could be used to measure the direction and the amplitude of the applied strain, at a given twist angle, or to correct the latter to bring it closer to the MA.
On the other hand, we showed that strain could be tuned to flatten the lowest energy bands in TBLG which may stabilize a superconducting state over a wide range of twist angles above the MA. 
We have also explained the absence of MA in t-BTMD and shown that the width of the lowest-energy bands is not so sensitive to strain as TBLG.
Our continuum models could be extended to investigate twisted van der Waals heterostructures \cite{Falko19}. These issues are left for further investigations. 

%%%%%%%%%%%%%%%%%%%%%%%%%%%%%%%%%%%%%%%%%%%%%%%%%%%%%%%%%%%%%%%%%%%%%%%%%%%%%%%%%%%%%%%%
\section{Acknowledgment}
This work was supported by the Tunisian Ministry of Higher Education and Scientific Research.
The authors acknowledges the kind hospitality of ICTP (Trieste, Italy), where the work was started. S.H. was supported by a Simons-ICTP associate fellowship.

\vspace{1cm}
$^{\ast}$ Electronic address: sonia.haddad@fst.utm.tn

\newpage
\widetext
\begin{center}
\textbf{\large Supplemental material for twistronics versus straintronics in twisted bilayers of graphene and transition metal dichalcogenides}
\end{center}
 
\section{Derivation of the low-energy Hamiltonian of strained TBLG}\label{TBLG}

We consider, as in Ref.\onlinecite{Bi}, that the two layers are rotated and strained oppositely to preserve the orientation of the moiré Brillouin zone \cite{Jung}.
We focus on heterostrain since homostrain, where both layer are subject to identical strain, is found to slightly affect the moiré structure \cite{Guy}.\

To derive the low-energy Hamiltonian of unstrained TBLG, Bistritzer and MacDonald \cite{Mc11} considered the leading terms in the interlayer tunneling amplitudes, which reduce to three nearest hopping processes in momentum space connecting states $|\mathbf {k}\rangle_1$ around the Dirac point $\mathbf{D}_{1,\xi}$ of layer $1$, to the states $|\mathbf{k+q_{j\xi}}\rangle_2$ around $\mathbf{D}_{2,\xi}$, the Dirac point of layer $2$. The $\mathbf{q_{j\xi}}$ vectors are given by
\begin{eqnarray}
 &&\mathbf{q}_{1\xi}=\xi k_{\theta}\left(0,1\right),\; 
 \mathbf{q}_{2\xi}=\mathbf{q}_{1\xi}+\xi\mathbf{G}^M_1=
 \xi k_{\theta}\left(-\frac{\sqrt{3}}2,-\frac12\right),\nonumber\\
 &&\mathbf{q_3}=\mathbf{q}_{1\xi}+\xi\left(\mathbf{G}^M_1+\mathbf{G}^M_2\right)=
 \xi k_{\theta}\left(\frac{\sqrt{3}}2,-\frac12\right),\nonumber\\
 \label{q0}
\end{eqnarray}
where $k_{\theta}=2k_D\sin\frac{\theta}2\sim \theta k_D$ and $k_D=|\mathbf{k}_{D1,\xi}|=|\mathbf{k}_{D2,\xi}|=\frac {4\pi}{3a}$, $a$ being the graphene lattice parameter. The $\left(\mathbf{G}^M_1,\mathbf{G}^M_2\right)$ is the moiré BZ basis (Fig.1 (g) of the main text) given by \cite{Bi}: $\mathbf{G}^M_i=\mathcal{E}_t^T\mathbf{G}_i$, $\mathcal{E}_t$ being the total deformation tensor, $\mathbf{G}_i$ are the lattice basis vectors of the monolayer reciprocal lattice $\mathbf{G}_1=\frac{2\pi}a \left(1,-1/\sqrt{3}\right)$ and $\mathbf{G}_2=\frac{2\pi}a \left(0,2/\sqrt{3}\right)$. \
The Dirac point of a layer satisfies $\mathbf{D}_{1,\xi}=-\xi \frac 13\left(2\mathbf{G}_1 +\mathbf{G}_2\right)$, which leads to $\sum^3_{j=1} \mathbf{q}_{j\xi}=\mathbf{0}$ for the unstrained TBLG.\

In the basis $\left\{|\mathbf {k}\rangle_1,|\mathbf{k+q_{j,\xi}}\rangle_2 \right\}$, the Hamiltonian, at the valley $\xi$, reads as \cite{Mc11}
\begin{eqnarray}
H(\mathbf{k})= 
\begin{pmatrix}
h_1(\mathbf{k}) & T_1 & T_2 & T_3\\
T^{\dagger}_1 & h_{2}(\mathbf{k+q_{1,\xi}}) & 0&0\\
T^{\dagger}_2 & 0& h_{2}(\mathbf{k+q_{2,\xi}}) & 0\\
T^{\dagger}_3 & 0& 0& h_{2}(\mathbf{k+q_{3,\xi}})\\
\end{pmatrix},\nonumber\\
\label{HBL0}
\end{eqnarray}
For the relaxed TBLG the $T_j$ matrices are given by \cite{Koshino18}
\begin{eqnarray}
T_1= 
\begin{pmatrix}
w & w^{\prime}\\
w^{\prime} & w^{\prime\prime}\
\end{pmatrix},
T_2= e^{i\xi\mathbf{G}^M_1\cdot\mathbf{r}}
\begin{pmatrix}
w & w^{\prime}e^{-i\xi\Phi}\\
w^{\prime}e^{i\xi\Phi}& w^{\prime\prime}\\
\end{pmatrix},
T_3= e^{i\xi\left(\mathbf{G}^M_1+\mathbf{G}^M_2\right)\cdot\mathbf{r}}
\begin{pmatrix}
w & w^{\prime}e^{i\xi\Phi}\\
w^{\prime}e^{-i\xi\Phi}& w^{\prime\prime}\
\end{pmatrix},
\label{T}
\end{eqnarray}
where $\Phi=\frac{2\pi}3$, $\mathbf{r}$ is the shortest inplane shifts between carbon atoms of the two layers \cite{Falko19}. In the AA stacking configuration, each A atom of the top layer is directly located above an A atom of the bottom layer, while for AB stacking, the A atom of the top layer has a B atom of the bottom layer whereas the top B atom has no partner \cite{MacDo-Rev}. For the AA stacking regions, $\mathbf{r}= \mathbf{0}$ while $\mathbf{r}= \pm \frac{a}{\sqrt{3}}\mathbf{e}_x$ for respectively the AB and BA-stacked regions \cite{Falko19}.
We focus on the AA stacking domains showing peaked LDOS and high electric conductivity \cite{Gadelha,Zhang}.\

In Eq. [\ref{T}] $w$, $w^{\prime}$ and $w^{\prime\prime}$ are the tunneling amplitudes which are equal $w=w^{\prime}=w^{\prime\prime}\sim 118 \mathrm{meV}$ in the rigid TBLG  \cite{Bi}.
The role of lattice relaxation will be discussed in the next section.\
 
In Eq.\ref{HBL0}, $h_1(\mathbf{k})=-\hbar v_F\left(\xi\sigma_x,\sigma_y\right)\cdot\mathbf{k} $ and $h_2(\mathbf{k}_j)=h_2(\mathbf{k}+\mathbf{q}_{j\xi})$, where the momentum $\mathbf{k}$ is written relatively to $\mathbf{D}_{1,\xi}$.\

The $\mathbf{k}$ dependent term in Eq. [\ref{HBL0}] is treated as a perturbation and to the leading order in $\mathbf{k}$, the effective Hamiltonian can be written as
\begin{eqnarray}
 H^{(1)}\left(\mathbf{k}\right)=\frac{\langle\Psi|H(\mathbf{k})|\Psi\rangle}{\langle\Psi|\Psi\rangle},
 \label{H10}
\end{eqnarray}
where $\Psi=\left(\psi_0(\mathbf{k}),\psi_1(\mathbf{k}),\psi_2(\mathbf{k}),\psi_3(\mathbf{k})\right)$ is the zero energy eigenstate of $H\left(\mathbf{k}=\mathbf{0}\right)$.  $\Psi$ is constructed on the two-component sublattice spinor $\psi_0(\mathbf{k})$ ($\psi_j(\mathbf{k})$) of layer $1$ (layer $2$) taken at the momentum $\mathbf{k}$ ($\mathbf{k}+\mathbf{q}_{j\xi}$) around the Dirac point $\mathbf{D}_{1,\xi}$ ($\mathbf{D}_{2,\xi}$) at a given valley $\xi$.
The $\Psi$ components satisfy
\begin{eqnarray}
 h_1\psi_0+\sum_j T_j\psi_j=0,\,\mathrm{and}\, T^{\dagger}_j\psi_0+h_j\psi_j=0,
\end{eqnarray}
where $h_j\equiv h_2(\mathbf{q}_{j,\xi})$.
Since $\psi_0$ is the zero energy eigenstate of $h_1$, then
\begin{eqnarray}
 \psi_j=-h^{-1}_j T^{\dagger}_j\psi_0, \mathrm{and}\; \sum_jT_jh^{-1}_j T^{\dagger}_j=0.
 \label{psij}
\end{eqnarray}
To the leading order in $\mathbf{k}$, and neglecting $\theta$ in $h_j$ and $h_1$ \cite{Mc11},  $H^{(1)}\left(\mathbf{k}\right)$, takes the following form
\begin{eqnarray}
 H^{(1)}\left(\mathbf{k}\right)=\frac{\langle\Psi|H(\mathbf{k})|\Psi\rangle}{\langle\Psi|\Psi\rangle}=\frac1{\langle\Psi|\Psi\rangle}
 \left[\psi^{\dagger}_0h_1\left(\mathbf{k}\right) \psi_0+ \psi^{\dagger}_0\sum_jT_j h^{-1}_j h_1\left(\mathbf{k}\right)h^{-1}_j T^{\dagger}_j \psi_0\right]=-\hbar \psi^{\dagger}_0 v^{\ast} \mathbf{\sigma^{\ast}}\cdot \mathbf{k}\psi_0,
 \label{psiHpsi}
\end{eqnarray}
 where $v^{\ast}=v_F\frac{1-3\alpha^2}{1+6\alpha^2}$ and $\sigma^{\ast}=\left(\xi\sigma_x,\sigma_y\right)$.\
 
 It is worth stressing that the continuum model cannot catch the stacking dependence of the low-energy Hamiltonian, since the $\mathbf{r}$ dependence of the $T_j$ matrices cancel out in Eq.\ref{psiHpsi}, which is not consistent with the tight binding calculations showing peaked LDOS at the AA regions \cite{Gadelha,Zhang}.\
 
In the strained TBLG, the Hamiltonian of layer $(i)$ rotated at a small angle $\theta _i$ and subject to a strain tensor $\epsilon_i$ is written as
\begin{eqnarray}
 h_1\left(\mathbf{k}\right)=-\hbar v_F \mathbf{\sigma^{\ast}}\cdot \left(\mathbb{I}+ \mathcal{E}_{t1}-\beta\mathcal{E}_1\right) \mathbf{k}\nonumber\\
  h_2\left(\mathbf{k}\right)=-\hbar v_F \mathbf{\sigma^{\ast}}\cdot \left(\mathbb{I}+ \mathcal{E}_{t2}-\beta\mathcal{E}_2\right) \left(\mathbf{k}+\mathbf{q}_{j\xi}\right)
  \label{hj}
\end{eqnarray}

where $\mathcal{E}_{ti}=\mathcal{E}_i+R(\theta_i)$ is the total deformation tensor of layer $(i)$ and $R(\theta_i)$ is the rotation matrix.\

Now, the $\mathbf{q}_{j,\xi}$ vectors depend on the displacement of the Dirac points under strain as:
\begin{eqnarray}
 \mathbf{q}_{1,\xi}=\mathbf{k}_{D1,\xi}-\mathbf{k}_{D2,\xi}=\mathcal{E}^T_{t}\mathbf{K}^{\xi}_{0}+\xi\mathbf{A} , \; 
 \mathbf{q}_{2,\xi}=\mathbf{q}_{1,\xi}+\xi\mathbf{G}^M_1, \mathbf{q}_{3,\xi}=\mathbf{q}_{1\xi}+\xi\left(\mathbf{G}^M_1+\mathbf{G}^M_2\right)
\end{eqnarray}
As in Ref.\onlinecite{Mc11}, the unperturbed Hamiltonian, corresponds to $\mathbf{k}=\mathbf{0}$, then the $h_j$ terms read as
$h_j=-\hbar v_F\mathbf{\sigma^{\ast}}\cdot\mathbf{Q}_{j,\xi}$ where 
\begin{eqnarray}
 \mathbf{Q}_{j,\xi}=\left(\mathbb{I}+ \mathcal{E}_{t}-\beta\epsilon\right) \mathbf{q_{j\xi}}
 \label{Qj}
\end{eqnarray}
The interlayer tunneling coupling $w$ (Eq. [\ref{T}]) is found to be unchanged to the first order in strain amplitude \cite{Falko19}.  We, then, consider strain independent tunneling amplitudes for the rigid and the relaxed TBLG. \

To the first order in strain, we set, in Eq. [\ref{Qj}], $\mathbf{Q}_{j,\xi}\sim\mathbf{q}_{j,\xi}$ and $\mathbf{q}_{j,\xi}=\mathbf{q}^0_{j,\xi}+\Delta\mathbf{q}_{j,\xi} $, where $\mathbf{q}^0_{j\xi}$ correspond to vectors of the unstrained TBLG given by Eq. [\ref{q0}]. This approximation yields to $2|\epsilon_{ij}|+\theta\ll 1$. \newline
For $\epsilon_{ij}\sim 1\%$ and taking an error of $2|\epsilon_{ij}|+\theta \sim 5\%$, gives $\theta \sim 1.7^{\circ}$ for TBLG. We, then, expect that our analytical expressions will be in agreement with tight binding and first principles calculations for twist angles not exceeding $2^{\circ}$.
This point will be discussed in the next session, for the relaxed TBLG.\

The amplitude of the four two-component spinor $\Psi$ is $\langle\Psi|\Psi\rangle=|\psi_0|^2+\sum_j|\psi_j|^2$ with $\psi_j$ given by Eq. [\ref{psij}], where $h^{-1}_j =-\frac1{\hbar v_F|\mathbf{q}_{j,\xi}|^2 }\mathbf{\sigma^{\ast}}\cdot\mathbf{q}_j$. 
To the leading term in strain, $\Psi$ amplitude is unchanged compared with the unstrained case $\langle\Psi|\Psi\rangle\sim 1+6\alpha^2$.\

To the first order in $\mathbf{k}$, the effective Hamiltonian becomes
\begin{eqnarray}
 H^{(1)}\left(\mathbf{k}\right)=\frac{\langle\Psi|H(\mathbf{k})|\Psi\rangle}{\langle\Psi|\Psi\rangle}=\frac1{\langle\Psi|\Psi\rangle}
 \left[\psi^{\dagger}_0h_0\left(\mathbf{\tilde{k}}\right) \psi_0+ \psi^{\dagger}_0\sum_jT_j h^{-1}_j h_0\left(\mathbf{\tilde{k}}\right)h^{-1}_j T^{\dagger}_j \psi_0\right],
\end{eqnarray}
where $\mathbf{\tilde{k}}=\left(\mathbb{I}+ \mathcal{E}_{t}-\beta\epsilon\right) \mathbf{k}$.\

To the lowest order in strain, the Hamiltonian $H^{(1)}$ reduce to Eq.7, where the tilt parameter $\mathbf{v}_0=\left(v_{0x},v_{0y}\right)$ and renormalized velocities are given by 
 \begin{eqnarray}
 &&v_{0x}=2v_F\frac{\alpha^2}{k_{\theta}}\left[ 2\Delta q_{1y,\xi}-\Delta q_{2y,\xi}-\Delta q_{2y,\xi}+\sqrt{3}\left( \Delta q_{2x,\xi}-\Delta q_{3x\xi}\right)\right],\nonumber\\
&&v_{0y}=2v_F\frac{\alpha^2}{k_{\theta}}\left[ 2\Delta q_{1x,\xi}-\Delta q_{2x,\xi}-\Delta q_{3x,\xi}-\sqrt{3}\left( \Delta q_{2y,\xi}-\Delta q_{3y,\xi}\right)\right],\nonumber\\
&&v_{x}=v_F\left[1-3\alpha^2\left(1+\left(1-\beta\right)\epsilon_{xx}\right)\right]
+\frac{v_F\alpha^2}{k_{\theta}}\left[\Delta q_{3y,\xi}+\Delta q_{2y,\xi}+4\xi\Delta q_{1y,\xi}-\sqrt{3}\xi\left( \Delta q_{2x,\xi}-\Delta q_{3x,\xi}\right)\right],\nonumber\\
&&v_{y}=v_F\left[1-3\alpha^2\left(1+\left(1-\beta\right)\epsilon_{yy}\right)\right]
+\frac{v_F\alpha^2}{k_{\theta}}\left[-3\xi \left(\Delta q_{3y,\xi}+\Delta q_{2y,\xi}\right)-\sqrt{3}\xi\left( \Delta q_{2x,\xi}-\Delta q_{3x,\xi}\right)\right],\nonumber\\
&&v_{xy}=-3v_F \alpha^2\left[\theta+\left(1-\beta\right)\epsilon_{xy}\right]+
\frac{v_F\alpha^2}{k_{\theta}}\left[\xi\left(\Delta q_{3x,\xi}+\Delta q_{2x,\xi}\right)
+4\xi \Delta q_{1x}
+ \sqrt{3}\xi\left( \Delta q_{2y,\xi}-\Delta q_{3y,\xi}\right)\right],\nonumber\\
&&v_{yx}=-3v_F \alpha^2\left[\left(1-\beta\right)\epsilon_{xy}-\theta\right]+
\frac{v_F\alpha^2}{k_{\theta}}\left[3\xi\left(\Delta q_{3x,\xi}+\Delta q_{2x,\xi}\right)-\sqrt{3}\xi\left(\Delta q_{2y,\xi}-\Delta q_{3y,\xi}\right)\right].
 \label{v*2}
\end{eqnarray}
It is worth to note that regarding the small value of the twist angle, the terms $\left(1-\beta\right)\epsilon_{ij}, (i,j=x,y)$ in Eq. [\ref{v*2}] could be neglected compared with the corrections of the form $\frac{\Delta q{ji}}{k_{\theta}},i=x,y, j=1,2,3$.

The $\Delta q_{ji}$ are the components of the strain-induced corrections $\Delta\mathbf{q}_{j,\xi}$ to the hopping vectors $\mathbf{q}_{0j,\xi}$ of the unstrained TBLG and are given by
\begin{eqnarray}
 &&\Delta\mathbf{q}_{1\xi}=\xi\left(-\frac{4\pi}{3a}\epsilon_{xx}+A_x,-\frac{4\pi}{3a}\epsilon_{xy}+A_y\right),
 \mathrm{where} \,\mathbf{A}=\frac{\sqrt{3}}{2a}\beta\left( \epsilon_{xx}-\epsilon_{yy},-2\epsilon_{xy}\right)\nonumber\\
 && \Delta\mathbf{q}_{2\xi}=\left(\Delta q_{1x,\xi}+\xi\frac{2\pi}{a}\left(\epsilon_{xx}-\frac1{\sqrt{3}}\epsilon_{xy}\right),
  \Delta q_{1y,\xi}+\xi\frac{2\pi}{a}\left(\epsilon_{xy}-\frac1{\sqrt{3}}\epsilon_{yy}\right) \right),\nonumber\\
 && \Delta\mathbf{q}_{3\xi}=\left(\Delta q_{1x,\xi}+\xi\frac{2\pi}{a}\left(\epsilon_{xx}+\frac1{\sqrt{3}}\epsilon_{xy}\right),
  \Delta q_{1y,\xi}+\xi\frac{2\pi}{a}\left(\epsilon_{xy}+\frac1{\sqrt{3}}\epsilon_{yy}\right)\right),\nonumber\\ 
  \label{delatq}
\end{eqnarray}
At small angle, one can neglect, in Eq. [\ref{v*2}], the correction terms of the form $\epsilon_{ij}\left(1-\beta\right)$ and $\epsilon_{ij}\left(1-\beta\right)-\theta$  compared with $\frac{\Delta q_j}{k_{\theta}}\sim \frac{\epsilon_{ij}}{\theta} $.
The expressions given by Eq. [\ref{v*2}] lead to those of Eq.8 of the main text.\

Considering the valley $\xi=-$, as in Ref. \onlinecite{Bi}, $v_y$ ($v_x$) velocity component (Eq. [\ref{v*2}]) vanishes under a compressive (tensile) shear strain described by the tensor $\epsilon_{ii}=0$ and $\epsilon_{xy}=\epsilon_{yx}=\epsilon<0$. This type of strain, has been already realized in graphene monolayer \cite{shearML,shearML2}.

\section{Derivation of the low-energy Hamiltonian of relaxed TBLG under strain}\label{r-TBLG}

In the TBLG relaxed lattice, the interlayer tunneling parameters (Eq.\ref{T}) are no more equal and are given by $w\sim w^{\prime\prime} \sim 0.090$ eV and $w^{\prime}=0.117$ eV \cite{Sarma2,Koshino20}. 
For simplicity, we consider an AA-stacked bilayer where the $T_j$ matrices take the form:
\begin{eqnarray}
 T_1=w\left(\mathbb{I} +u \sigma_x\right), \, 
 T_2= w\left[\mathbb{I} +u \left(-\frac 12\sigma_x+\xi\frac{\sqrt{3}}2\sigma_y\right)\right],
 T_3= w\left[\mathbb{I} +u \left(-\frac 12\sigma_x-\xi\frac{\sqrt{3}}2\sigma_y\right)\right],
\end{eqnarray}
here $u=w^{\prime}/w$.\newline
Following the same procedure as the previous section, we derive the low-energy effective Hamiltonian of the relaxed TBLG under strain, which is given by
\begin{eqnarray}
 H^{(1)}\left(\mathbf{k}\right)=\frac{-\hbar}{\langle\Psi|\Psi\rangle}\psi^{\dagger}_0\left[v_{0x}^r k_x+ v_{0y}^r k_y+ \xi \sigma_x v_x^r k_x +
 \sigma_y v_y^r k_y+\xi \sigma_x v_{xy}^rk_y+\sigma_y v_{yx}^rk_x\right] \psi_0
 \label{Hrstrain}
\end{eqnarray}
where $\langle\Psi|\Psi\rangle=1+3\left(1+u^2\right)\alpha^2$ and the effective velocities are

 \begin{eqnarray}
v_{0x}^r&=&2v_F\frac{\alpha^2}{k_{\theta}}u\left[2\Delta q_{1y,\xi}-\Delta q_{2y,\xi}-\Delta q_{3y,\xi}+\sqrt{3}u\left( \Delta q_{2x,\xi}-\Delta q_{3x,\xi}\right)\right],\nonumber\\
v_{0y}&=&2v_F\frac{\alpha^2}{k_{\theta}}u\left[ 2\Delta q_{1x,\xi}-\Delta q_{3x,\xi}-\Delta q_{2x,\xi}-\sqrt{3}u\left( \Delta q_{2y,\xi}-\Delta q_{3y,\xi}\right)\right],\nonumber\\
v_{x}^r&=&v_F\left[1-3\alpha^2u^2 
+\frac{\alpha^2}{k_{\theta}}\left(\xi\left(2-u^2\right)\left(\Delta q_{3y,\xi}+\Delta q_{2y,\xi}\right)
+ 2\xi\left(1+u^2\right)\Delta q_{1y,\xi}-\sqrt{3}u^2\xi\left(\Delta q_{2x,\xi}-\Delta q_{3x,\xi}\right)\right)\right],\nonumber\\
v_{y}^r&=&v_F\left[1-3\alpha^2u^2
-\frac{\alpha^2}{k_{\theta}}\left(\xi\left(2+u^2\right)\left(\Delta q_{3y,\xi}+\Delta q_{2y,\xi}\right)
+\xi\sqrt{3}u^2\left(\Delta q_{2x,\xi}-\Delta q_{3x,\xi}\right)+2\xi\left(1-u^2\right)\Delta q_{1y}\right)\right],\nonumber\\
v_{xy}^r&=&
\frac{v_F\alpha^2}{k_{\theta}}\left[\xi\left(2-u^2\right)\left(\Delta q_{3x,\xi}+\Delta q_{2x,\xi}\right)+\xi\sqrt{3}u^2\left( \Delta q_{2y,\xi}-\Delta q_{3y,\xi}\right)+2\xi\left(1+u^2\right)\Delta q_{1x,\xi}\right],\nonumber\\
v_{yx}^r&=&
\frac{v_F\alpha^2}{k_{\theta}}\left[\xi\left(u^2+2\right)\left(\Delta q_{3x,\xi}+\Delta q_{2x,\xi}\right)-\xi\sqrt{3}u^2\left(\Delta q_{2y,\xi}-\Delta q_{3y,\xi}\right)+2\xi\left(1-u^2\right)\Delta q_{1x,\xi}\right],\nonumber\\
 \label{vrelax2}
\end{eqnarray}
For $u=1$, we recover the expressions of the rigid TBLG given by Eq. [\ref{v*2}] where we neglect the corrections of the form $\left(1-\beta\right)\epsilon_{ij}$ as discussed in the previous section.\

In the unstrained lattice, Eq. [\ref{v*2}] gives rise to the effective Fermi velocity
\begin{eqnarray}
v_r^{\ast}=v_F\frac{1-3u^2\alpha^2}{1+3\left(1+u^2\right)\alpha^2},
\label{vrelax3}
\end{eqnarray}
Ea. [\ref{vrelax3}] shows that the effective velocity of the unstrained TBLG is reduced by relaxation. This feature is in agreement with the first principles and TB calculations \cite{Koshino20,Guinea,Cantele} and with our numerical results carried out, at $\theta \sim 1.5^{\circ}$ within the continuum model and using a 128-state basis (Fig. 2 in the main text). However, for $\theta >1.7^{\circ}$, the width of the lowest energy bands of the relaxed TBLG are found to be larger compared with the rigid lattice \cite{Koshino20,Guinea,Cantele}. Eq. [\ref{vrelax3}] could not account for this behavior, which limits its range of applicability to angles around the MA not exceeding $\theta\sim 2^{\circ}$.
\section{Derivation of the low-energy Hamiltonian of twisted bilayer transition metal dichalcogenides under strain}\label{TMD}

We consider, as in the previous section, two TMD homolayers rotated and strained in opposite ways as done for TBLG (Eq. [\ref{HBL0}]) but now, one needs to replace the $h_i, (i=1, 2)$  by 
\begin{eqnarray}
 h_1^{\prime}\left(\mathbf{k}\right)&=&h_1\left(\mathbf{k}\right)+V_1\left(\mathbf{r}\right),\\
 h_2^{\prime}\left(\mathbf{k+q_{j,\xi}}\right)&=&h_2\left(\mathbf{k+q_{j,\xi}}\right)+V_2\left(\mathbf{r}\right)
\label{h12'}
 \end{eqnarray}
here $h_{1,2}$ are 
\begin{eqnarray}
h_1(\mathbf{k})&=&-\hbar v_F \left(\mathbb{I}+\mathcal{E}^T_t-\beta \mathcal{E}\right)\mathbf{k}
 \cdot \left(\xi \sigma_x,\sigma_y\right)
 +\frac m2 \left(\sigma_z+\mathbb{I}\right),\nonumber\\
h_{2,j}(\mathbf{k})&=&-\hbar v_F \left(\mathbb{I}+\mathcal{E}^T_{t2}-\beta \mathcal{E}_2\right) \left(\mathbf{k}+\mathbf{q}_{j\xi} \right)\cdot\left(\xi \sigma_x,\sigma_y\right)
+\frac m2 \left(\sigma_z+\mathbb{I}\right).
\label{h3TMD}
\end{eqnarray}
and the $V_l\left(\mathbf{r}\right), l=1,2$ is the intralayer potential written as \cite{Bi,Macdo18,Jung}
\begin{eqnarray}
 V_l\left(\mathbf{r}\right)=\sum_{i=1,2,3}
 \begin{pmatrix}
V_c e^{i\left(\mathbf{G}_i^M\cdot\mathbf{r}+(-1)^l \Phi_c\right)} & 0\\
0 & V_v e^{i\left(\mathbf{G}_i^M\cdot\mathbf{r}+(-1)^l \Phi_v\right)}\
\end{pmatrix}+h.c.,
\label{Vlayer}
\end{eqnarray}
where $V_c\sim 6.8 \mathrm{meV}$, $V_v\sim 8.9 \mathrm{meV}$, $\Phi_c\sim 89.7^{\circ}$ and $\Phi_v\sim 91.0^{\circ}$ for WSe$_2$ \cite{Macdo19,Bi,Jung}.\

In AA-stacked t-BTMD ($\mathbf{r}=\mathbf{0}$), $V_l$ is the same for both layers and reduces to
\begin{eqnarray}
 V_l\equiv V=
 \begin{pmatrix}
U_c  & 0\\
0 & U_v\
\end{pmatrix},
\label{V}
\end{eqnarray}
where $U_i=6V_i\cos\Phi_i, (i=v,c)$.\

The low-energy Hamiltonian of t-TMDs takes the same form as in Eq. [\ref{HBL0}] by replacing $h_1\left(\mathbf{k}\right)$ and $h_2\left(\mathbf{k}+\mathbf{q}_{j,\xi}\right)$ by $h^{\prime}_i\left(\mathbf{k}\right)=h_i\left(\mathbf{k}\right)+V, (i=1,2)$ where $h_i\left(\mathbf{k}\right)$ are given by Eq. [\ref{h3TMD}].
One can follow the same procedure, as in Ref. \onlinecite{Mc11} where the monolayer Hamiltonian $h_i$ is replaced by $h_i^{\prime}$ (Eq. [\ref{h12'}]), and $\psi_0$ is, now, the zero energy eigenstate of $h_1^{\prime}\left(\mathbf{k}\right)$. We, then, obtain:
\begin{eqnarray}
 H^{(1)}\left(\mathbf{k}\right)=\frac1{\langle\Psi|\Psi\rangle}
 \left[\psi^{\dagger}_0h_1^{\prime}\left(\mathbf{k}\right) \psi_0+ \psi^{\dagger}_0\sum_jT_j \left(h^{\prime}_j\right)^{-1}h_1^{\prime}\left(\mathbf{k}\right)\left(h^{\prime}_j\right)^{-1} T^{\dagger}_j \psi_0\right]
 \label{psiHpsi2}
\end{eqnarray}
To the leading order in strain and twist angle, $\left(h^{\prime}_j\right)^{-1}$ reads as
\begin{eqnarray}
 \left(h^{\prime}_j\right)^{-1}=\frac 1{X_j}\left[\hbar v_F \mathbf{\sigma^{\ast}}\cdot\mathbf{q}_{j,\xi}+
 a_+\mathbb{I} -a_-\sigma_z \right]
\end{eqnarray}
where $a_{\pm}=U_{\pm} +\frac m 2$, $U_{\pm}=\frac{U_c\pm U_v} 2$ (Eq. [\ref{V}]) and $X_j$ is
\begin{eqnarray}
 X_j=\gamma-\left(\hbar v_F\right)^2 \mathbf{q}_{j,\xi}^2
 \sim\left[\gamma-\left(\hbar v_F k_{\theta}\right)^2\right]
 \left[1-\frac{2 \left(\hbar v_F\right)^2 \mathbf{q}_{0j,\xi}.\Delta\mathbf{q}_{j\xi}}{\gamma-\left(\hbar v_F k_{\theta}\right)^2}\right],
 \label{Xj}
\end{eqnarray}
where $\gamma=\left(m+U_c\right)U_v$ and $\Delta\mathbf{q}_{j,\xi}$ are given by Eq. [\ref{delatq}] and
$h_1^{\prime}\left(\mathbf{k}\right)$ is written as
\begin{eqnarray}
 \left(h^{\prime}_1\right)=-\hbar v_F \mathbf{\sigma}^{\ast}\cdot\mathbf{k}+
 a_+\mathbb{I} +a_-\sigma_z
\end{eqnarray}
Following the same procedure as in the previous section, we found that, to the leading order in $\mathbf{k}$, the effective Hamiltonian of Eq. [\ref{psiHpsi2}] becomes
\begin{eqnarray}
 H^{(1)}\left(\mathbf{k}\right)=-\frac{\hbar v_F}{\langle\Psi|\Psi\rangle}\psi^{\dagger}_0\left[v^{\prime}_{0x} k_x+v^{\prime}_{0y} k_y+ \xi \sigma_x v^{\prime}_x k_x +
 \sigma_y v^{\prime}_y k_y+\sigma_x v^{\prime}_{xy}k_y+\sigma_y v^{\prime}_{yx}k_x\right] \psi_0
 \label{HTMDstrain}
\end{eqnarray}
where the tilt $\mathbf{v}^{\prime}=\left(v^{\prime}_{0x},v^{\prime}_{0y}\right)$ and the effective velocities are given by
\begin{eqnarray}
 v^{\prime}_{0x}&=&\left(\hbar v_F k_{\theta}\right)^2\left(\frac w{X_0}\right)^2
 \frac 2{k_{\theta}}
 \left[-2\left(\Delta q_{1y,\xi}+\Delta q_{2y,\xi}+\Delta q_{3y,\xi}\right)-Y \left(4\Delta q_{1y,\xi}+\Delta q_{2y,\xi}+\Delta q_{3y,\xi}\right)+\sqrt{3}Y\left(\Delta q_{3x,\xi}-\Delta q_{2x,\xi}\right)
\right]\nonumber\\
 &+&4 a_+\hbar v_F \left(\frac w{X_0}\right)^2 \left[\left(\Delta q_{1x,\xi}+\Delta q_{2x,\xi}+\Delta q_{3x,\xi}\right)+3Y\left(\Delta q_{3x,\xi}+\Delta q_{2x,\xi}\right)
 +\sqrt{3}Y\left(\Delta q_{2y,\xi}-\Delta q_{3y,\xi}\right)\right]\nonumber\\
v^{\prime}_{0y}&=&\left(\hbar v_F k_{\theta}\right)^2\left(\frac w{X_0}\right)^2
 \frac 2{k_{\theta}}
 \left[2\left(\Delta q_{1x,\xi}+\Delta q_{2x,\xi}+\Delta q_{3x,\xi}\right)+ 3 Y \left(\Delta q_{2x,\xi}+\Delta q_{3x,\xi}\right)
 +\sqrt{3}Y \left(\Delta q_{2y,\xi}-\Delta q_{3y,\xi}\right)\right]\nonumber\\
 &+&4 a_+\hbar v_F \left(\frac w{X_0}\right)^2\left[\Delta q_{1y,\xi}+\Delta q_{2y,\xi}+\Delta q_{3y,\xi}+Y\left( 4\Delta q_{1y,\xi}+\Delta q_{2y,\xi}+\Delta q_{3y,\xi}\right)+ \sqrt{3}Y\left(\Delta q_{2x,\xi}-\Delta q_{3x,\xi}\right)\right]\nonumber\\
v^{\prime}_{x}&=&\left(\hbar v_F k_{\theta}\right)^2\left(\frac w{X_0}\right)^2
\left[-3+\frac 2{k_{\theta}}\xi\left(\Delta q_{2y,\xi}+\Delta q_{3y,\xi}-2\Delta q_{1y,\xi}\right)
+\xi\frac{Y}{k_{\theta}} \left(\Delta q_{2y,\xi}+\Delta q_{3y,\xi}-8\Delta q_{1y,\xi}\right)+\sqrt{3}\xi \frac{Y}{k_{\theta}}\left(\Delta q_{2x,\xi}-\Delta q_{3x,\xi}\right)\right]\nonumber\\
 &+&2 a_+\hbar v_F \left(\frac w{X_0}\right)^2\xi\left(2\Delta q_{1x,\xi}-\Delta q_{2x,\xi}-\Delta q_{3x,\xi}-3Y\left(\Delta q_{2x,\xi}+\Delta q_{3x,\xi}\right)+\sqrt{3}Y
 \left(\Delta q_{3y,\xi}-\Delta q_{2y,\xi}\right)\right)\nonumber\\
 &+&\gamma\left(\frac w{X_0}\right)^2\left[3+2\xi\frac{Y}{k_{\theta}} \left(2\Delta q_{1y,\xi}+\sqrt{3}\left(\Delta q_{3x,\xi}-\Delta q_{2x,\xi}\right)-\left(\Delta q_{2y,\xi}+\Delta q_{3y,\xi}\right)\right)\right]\nonumber\\
 v^{\prime}_{y}&=&\left(\hbar v_F k_{\theta}\right)^2\left(\frac w{X_0}\right)^2
\left[-3+\xi\frac {2\sqrt{3}}{k_{\theta}}\left(\Delta q_{2x,\xi}-2\Delta q_{3x,\xi}\right)
+\xi Y\frac{3\sqrt{3}}{k_{\theta}} \left(\Delta q_{2x,\xi}-\Delta q_{3x,\xi}\right)
 +3\xi  \frac Y{k_{\theta}}\left(\Delta q_{2y,\xi}+\Delta q_{3y,\xi}\right)\right]\nonumber\\
 &+&2\xi \sqrt{3}a_+\hbar v_F \left(\frac w{X_0}\right)^2
 \left[\left(\Delta q_{2y,\xi}-\Delta q_{3y,\xi}\right)
 +\sqrt{3}Y\left(\Delta q_{2x,\xi}+\Delta q_{3x,\xi}\right)+Y\left(\Delta q_{2y,\xi}-\Delta q_{3y,\xi}\right)\right]\nonumber\\
 &+&\gamma\left(\frac w{X_0}\right)^2\left[3+2\xi\frac{Y}{k_{\theta}} \left(2\Delta q_{1y,\xi}+\sqrt{3}\left(\Delta q_{3x,\xi}-\Delta q_{2x,\xi}\right)-\left(\Delta q_{2y,\xi}+\Delta q_{3y,\xi}\right)\right)\right]\nonumber\\
 v^{\prime}_{xy}&=&\left(\hbar v_F k_{\theta}\right)^2\left(\frac w{X_0}\right)^2
\frac 1{k_{\theta}}\left[2\left(2\Delta q_{1x,\xi}-\Delta q_{2x,\xi}-\Delta q_{3x,\xi}\right)
-3Y\left(\Delta q_{2x,\xi}+\Delta q_{3x,\xi}\right)
 +\sqrt{3}Y\left(\Delta q_{3y,\xi}-\Delta q_{2y,\xi}\right)\right]\nonumber\\
 &+&2 a_+\hbar v_F k_{\theta} \left(\frac w{X_0}\right)^2\left[3\xi+\frac 1{k_{\theta}}\left(2\Delta q_{1y\xi}-\Delta q_{2y,\xi}-\Delta q_{3y\xi}\right)
 +\sqrt{3}\frac Y{k_{\theta}}\left(\Delta q_{3x,\xi}-\Delta q_{2x}\right)
 +\frac Y{k_{\theta}}\left(8\Delta q_{1y\xi}-\Delta q_{2y,\xi}-\Delta q_{3y\xi}\right)\right]\nonumber\\
v^{\prime}_{yx}&=&\left(\hbar v_F k_{\theta}\right)^2\left(\frac w{X_0}\right)^2
\frac 1{k_{\theta}}\left[2\sqrt{3}\xi\left(\Delta q_{3y,\xi}-\Delta q_{2y,\xi}\right)
-3Y\xi\left(\Delta q_{2x,\xi}+\Delta q_{3x,\xi}\right)
 -\xi \sqrt{3}Y\left(\Delta q_{3y,\xi}+\Delta q_{2y,\xi}\right)\right]\nonumber\\
 &+&2 \sqrt{3}a_+\hbar v_F k_{\theta}\left(\frac w{X_0}\right)^2\left[-\sqrt{3}+\xi\frac 1{k_{\theta}}\left(\Delta q_{2x,\xi}-\Delta q_{3x}\right)
 +3\xi\frac{Y}{k_{\theta}}\left(\Delta q_{2x,\xi}-\Delta q_{3x,\xi}\right)+\sqrt{3}\xi\frac{Y}{k_{\theta}} \left(\Delta q_{2y,\xi}+\Delta q_{3y,\xi}\right)\right],
 \label{v*TMD}
\end{eqnarray}
here $X_0=\gamma-\left(\hbar v_F k_{\theta}\right)^2$, $Y=\frac{\left(\hbar v_F k_{\theta}\right)^2}{X_0}$ and the amplitude of the two-component spinor is 
\begin{eqnarray}
 \langle\Psi|\Psi\rangle=|\Psi_0|^2 \left\{1+ 6\frac{\left(w\hbar v_F k_{\theta}\right)^2}
 {\left(|\gamma|+\left(\hbar v_F k_{\theta}\right)^2\right)^2}\left(1+\frac{a_+^2+a_-^2}{\left(\hbar v_F k_{\theta}\right)^2}\right)\right\}
 \label{PsiTMD}
\end{eqnarray}
In the absence of strain, the effective velocity $v^{\prime}_{x}$ and $v^{\prime}_{y}$ reduce to \begin{eqnarray}
 v^{\ast}_{TMD}= \frac{v_F}{\langle\Psi|\Psi\rangle} \left(1+\frac{3w^2}{\gamma-\left(\hbar v_Fk_{\theta}\right)^2}\right),
 \label{v_TMD2}
\end{eqnarray}
The latter yields to the renormalized velocity of undeformed TBLG for $m=0$ and $a_+=a_-=0$.\
We consider, as in Ref. \onlinecite{Bi}, the case of rigid twisted bilayer WSe$_2$ for which $w=9.7$ meV, 
$\hbar v_F/a\sim 1.1$ eV, $U_c=0.21$ meV, $U_v=-0.93$ meV and $\gamma=-1.16 10^{-3}\mathrm{(eV)^2}$ \cite{Bi,Jung}. In this case, and for a twist angle $\theta \ge 1^{\circ}$, 
$\left(\hbar v_F k_{\theta}\right)^2 \gg|\gamma|$, $|\Psi|^2$ becomes
\begin{eqnarray} 
 \langle\Psi|\Psi\rangle=|\Psi_0|^2 \left[1+ 6\alpha^2
\left(1+\frac{a_+^2+a_-^2}{\left(\hbar v_F k_{\theta}\right)^2}\right)\right]
\end{eqnarray}
with $\alpha=w/\hbar v_F k_{\theta}$
Regarding the small values of $w$ and $\gamma$, $v^{\ast}_{TMD}$ (Eq.[\ref{v_TMD2}]) could not vanish at any twist angle. However, flat bands can emerge below a critical angle $\theta_c$ for which the bandwidth could be considered narrow. This result is consistent with numerical calculations \cite{Naik18,Jung} reporting the occurrence of flat bands in t-BTMDs at a wide range of angles, where Fermi velocity does not vanish, but the lowest energy bands have narrow bandwidths. \

\begin{figure}[hpbt]
\begin{center}
$\begin{array}{cccc}
\includegraphics[width=0.4\columnwidth]{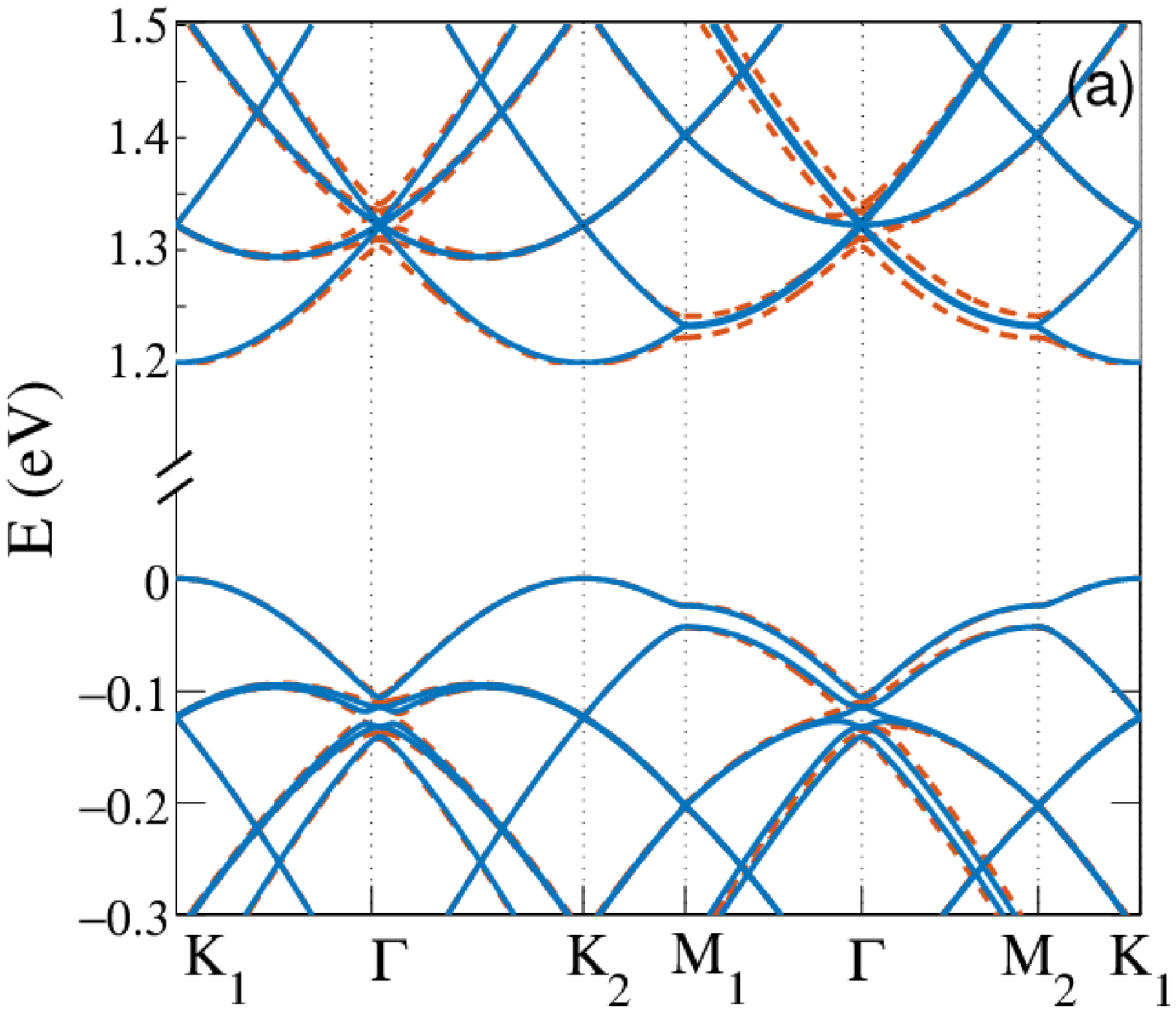}&
\includegraphics[width=0.4\columnwidth]{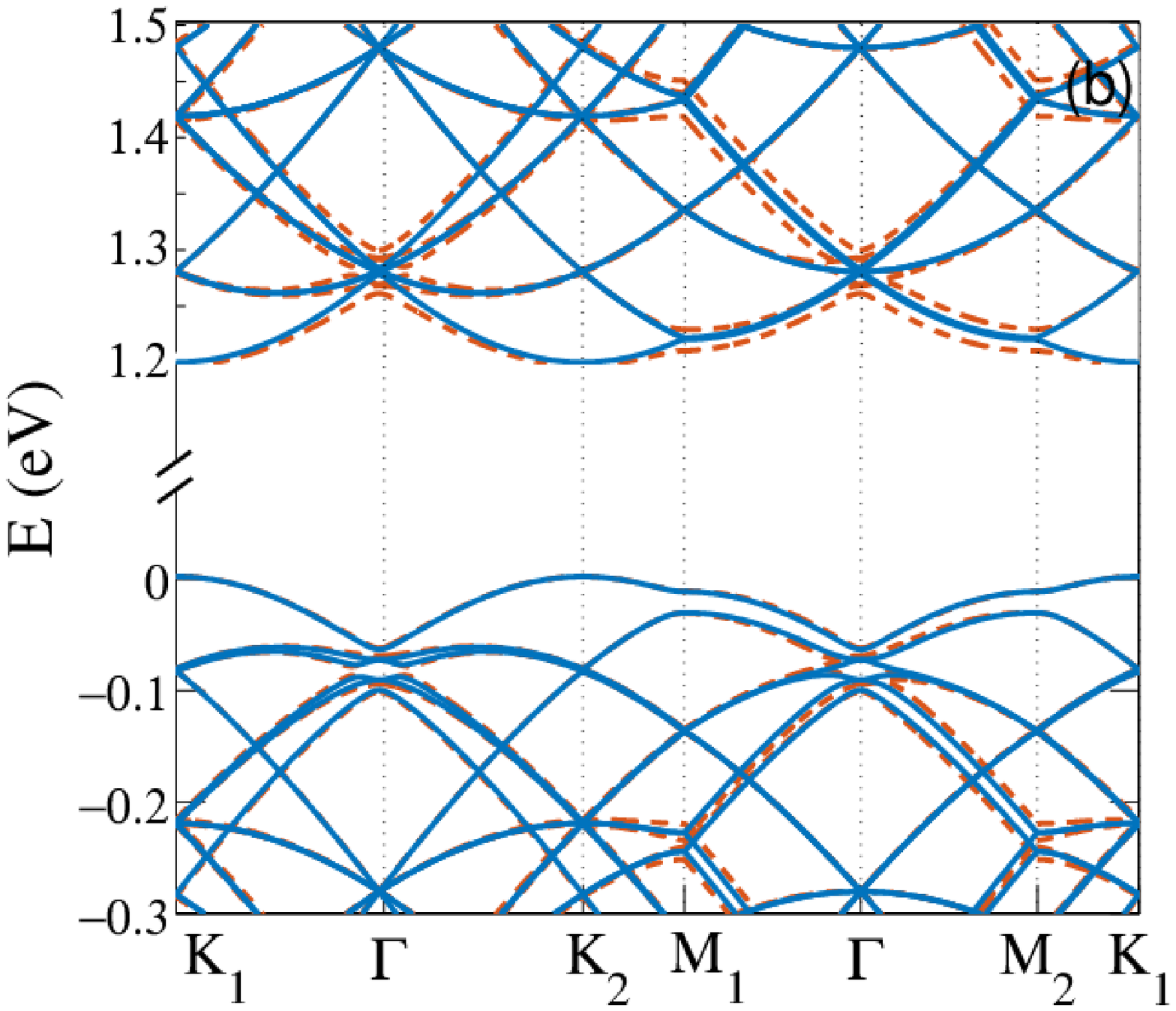}
\end{array}$
$\begin{array}{cccc}
\includegraphics[width=0.4\columnwidth]{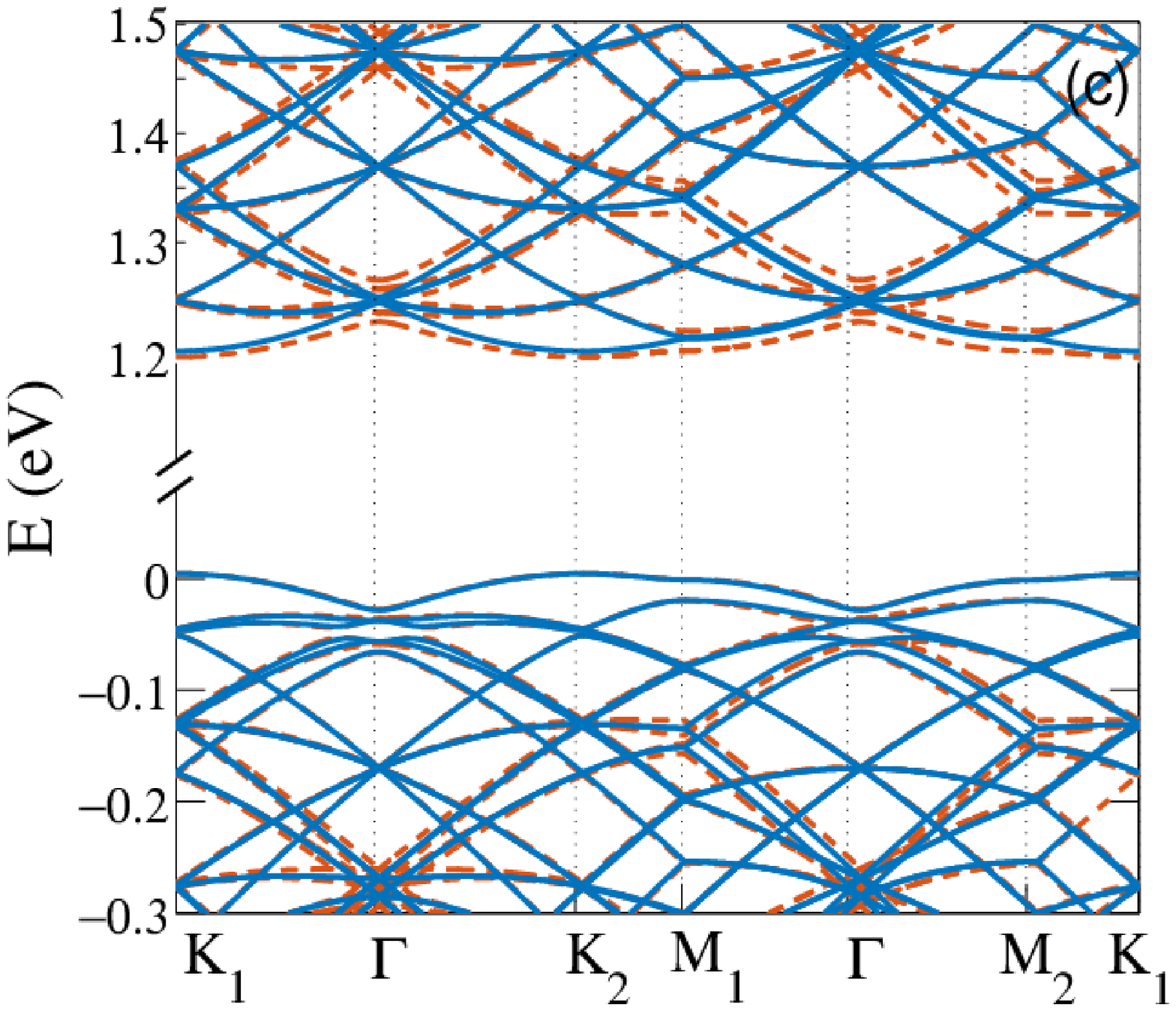}&
\includegraphics[width=0.4\columnwidth]{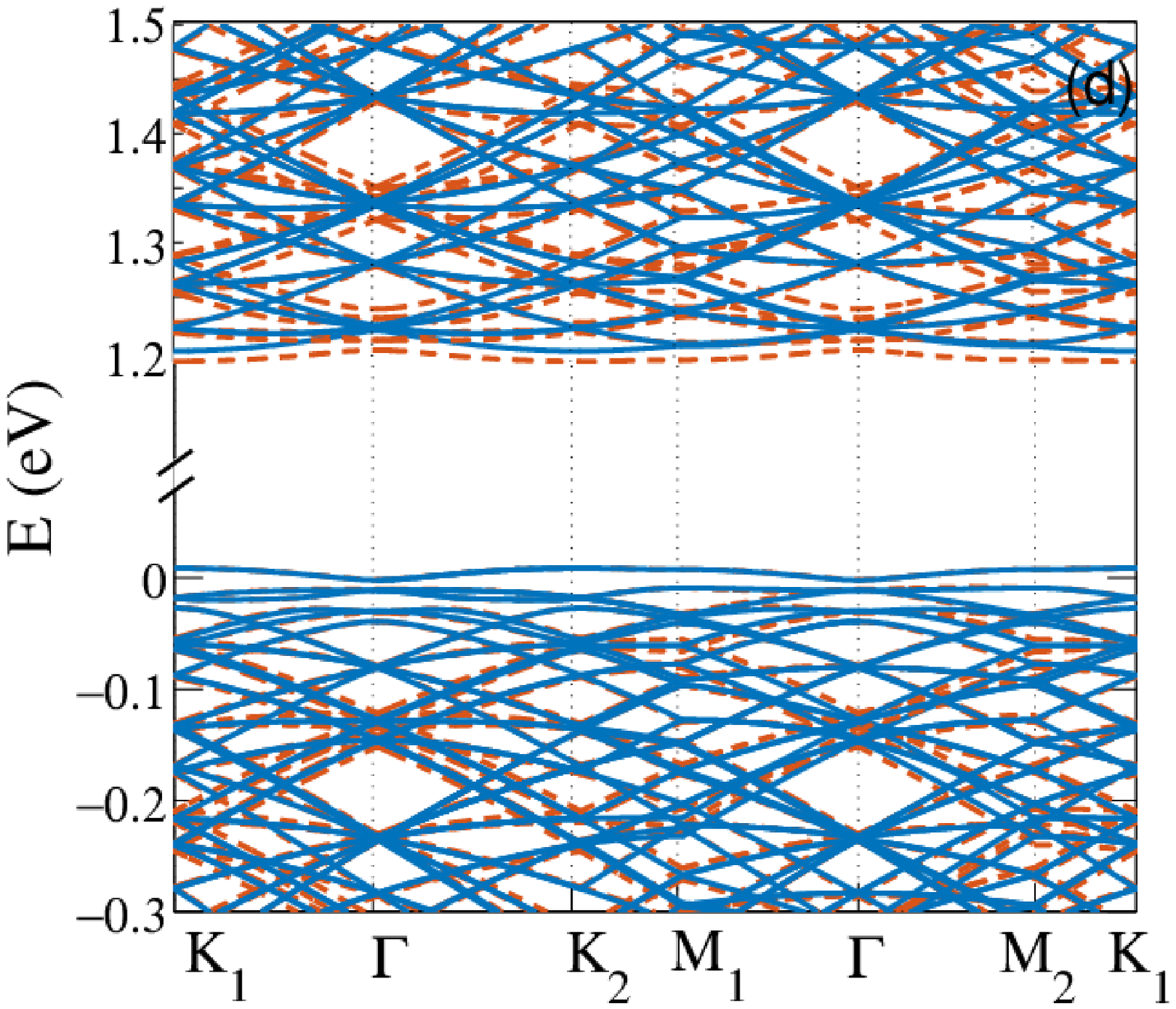}
\end{array}$
\end{center}
\caption{Low-energy band structure of unstrained WSe$2$ for the rigid (dashed lines) and relaxed lattice (solid lines) at twist angles  $\theta=5^{\circ}$ (a), $\theta=4^{\circ}$ (b), $\theta=3^{\circ}$ (c) and $\theta=2^{\circ}$ (d). The band structures are calculated by diagonalizing the low-energy Hamiltonian of t-TMBD \cite{supp} in the basis of $\left\{|\mathbf {k}\rangle_1,|\mathbf{k+q_j}\rangle_2\right\}$ constructed by the states around the Dirac point $\mathbf{D}_{1}$ of layer $1$ and those in the vicinity of the Dirac point $\mathbf{D}_{2}$ of layer $2$, respectively. 
Calculations are done for $\frac{\hbar v_F}a\sim 1.1$, $\beta=2.3$, $V_v=8.9$meV, $V_c=6.8$meV, $\Phi_v=91^{\circ}$, $\Phi_c=89.7^{\circ}$, $w=9.7$ meV for the rigid lattice and 
$w=1.1$ meV, $w^{\prime}=0$, and $w^{\prime\prime}=9.7$ meV for the relaxed system.}
\label{TMDfig1}
\end{figure}

Moreover, Fig. \ref{TMDfig1} shows that the band structure is, practically, not affected by the lattice relaxation, contrary to TBLG.

According to Eq. [\ref{v_TMD2}], magic angles with vanishing effective Fermi velocity may occur in t-TMDs if $\gamma$ is positive. For the same values of mass $m$, and intralayer potential parameters, $V_v$, $V_c$, $\Phi_c$ as WSe$_2$ but with $\Phi_v\le\frac{\pi}2$, $v^{\ast}_{TMD}$ (Eq. [\ref{v_TMD2}]) will vanish at a MA of $\theta_M\sim 0.5^{\circ}$.\

\begin{figure}[hpbt]
\begin{center}
$\begin{array}{cccc}
\includegraphics[width=0.4\columnwidth]{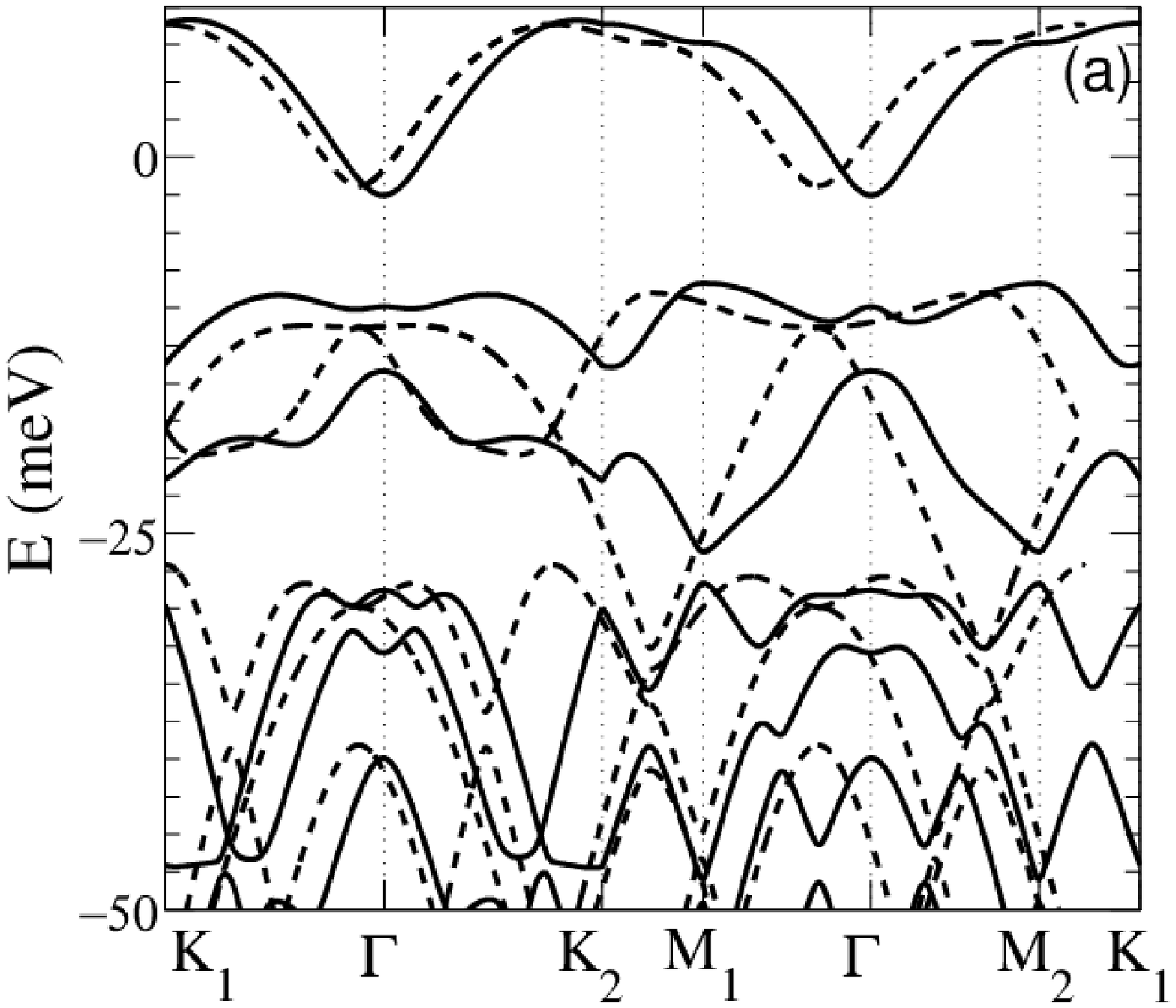}&
\includegraphics[width=0.4\columnwidth]{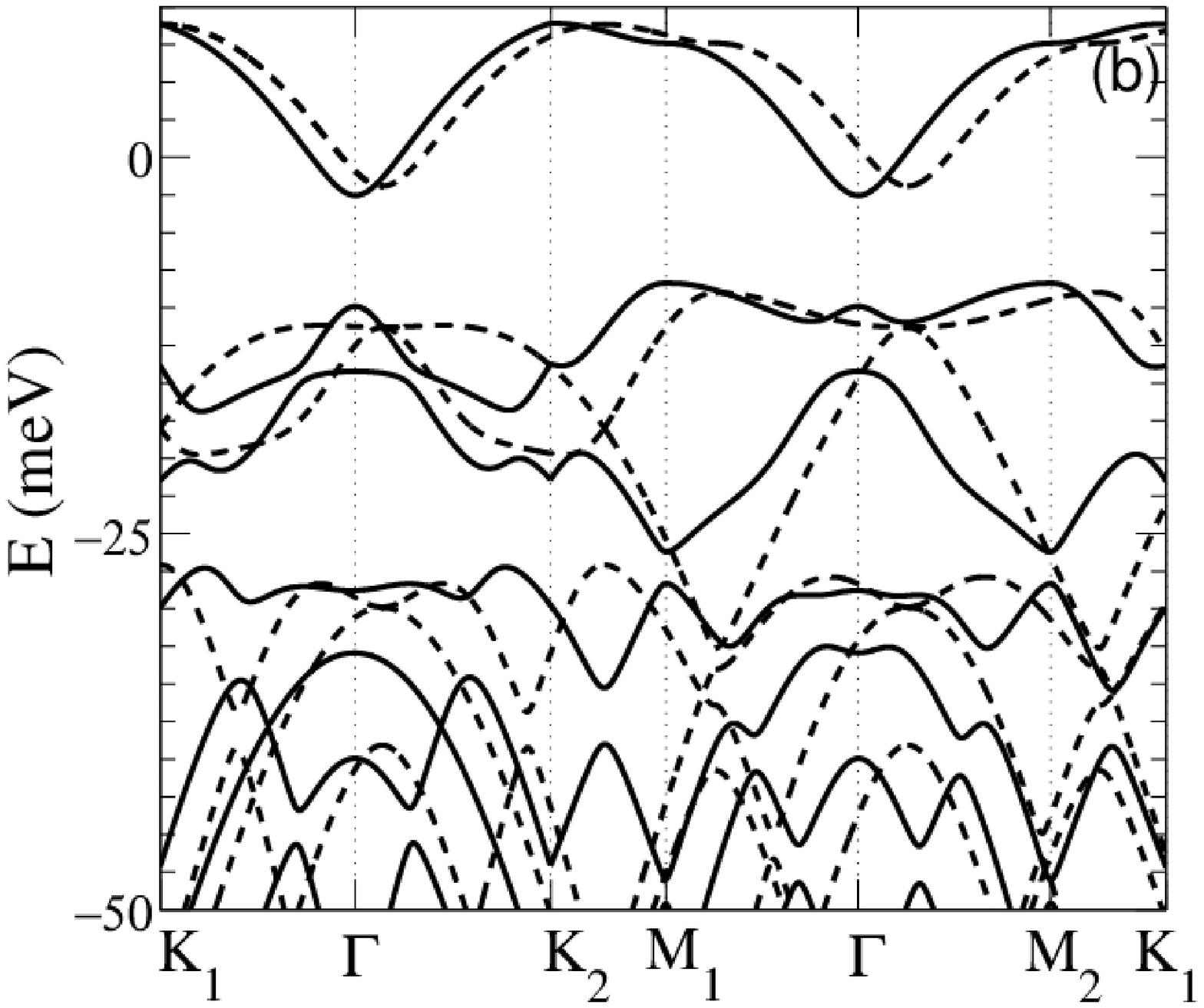}
\end{array}$
$\begin{array}{cccc}
\includegraphics[width=0.4\columnwidth]{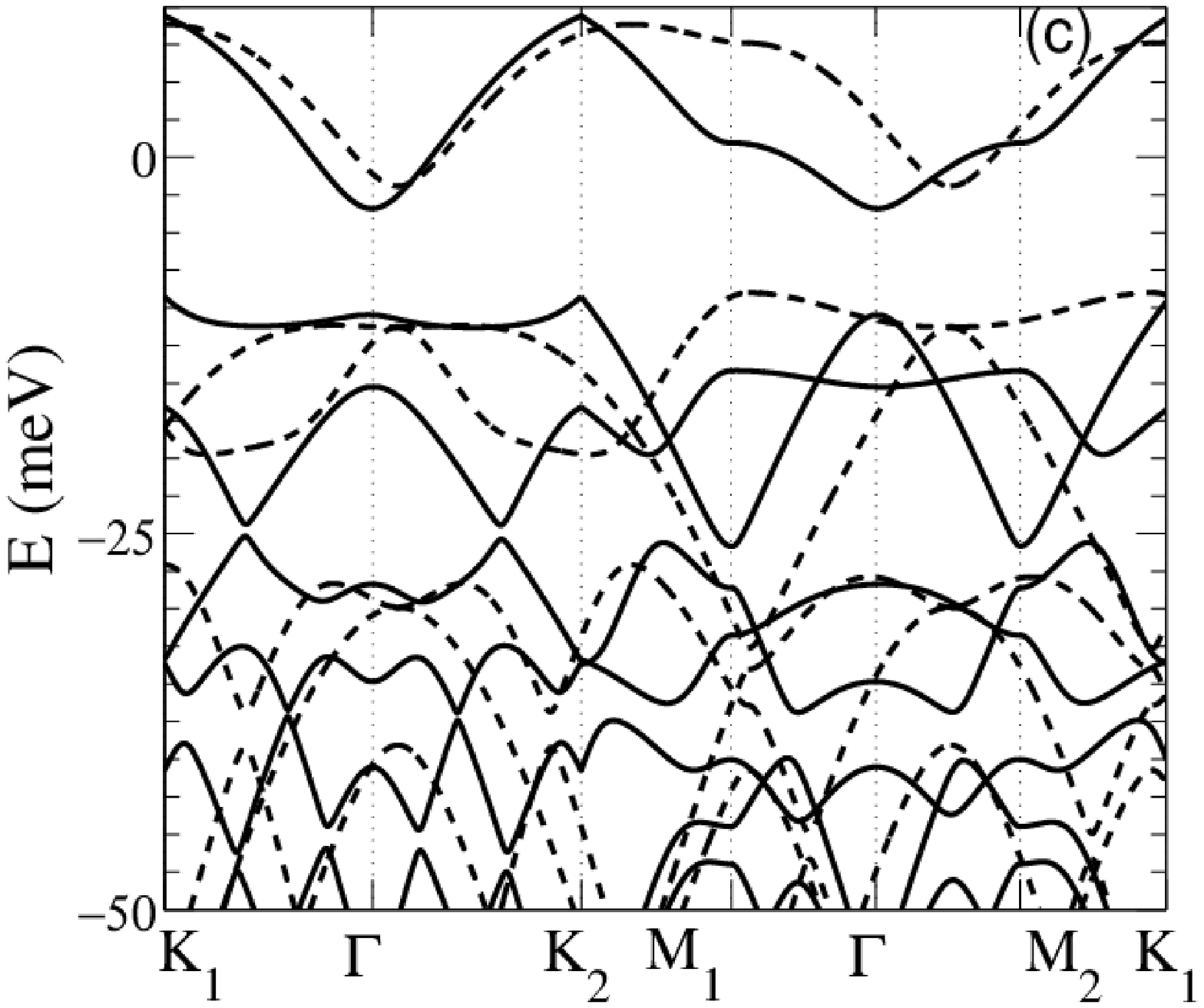}&
\includegraphics[width=0.4\columnwidth]{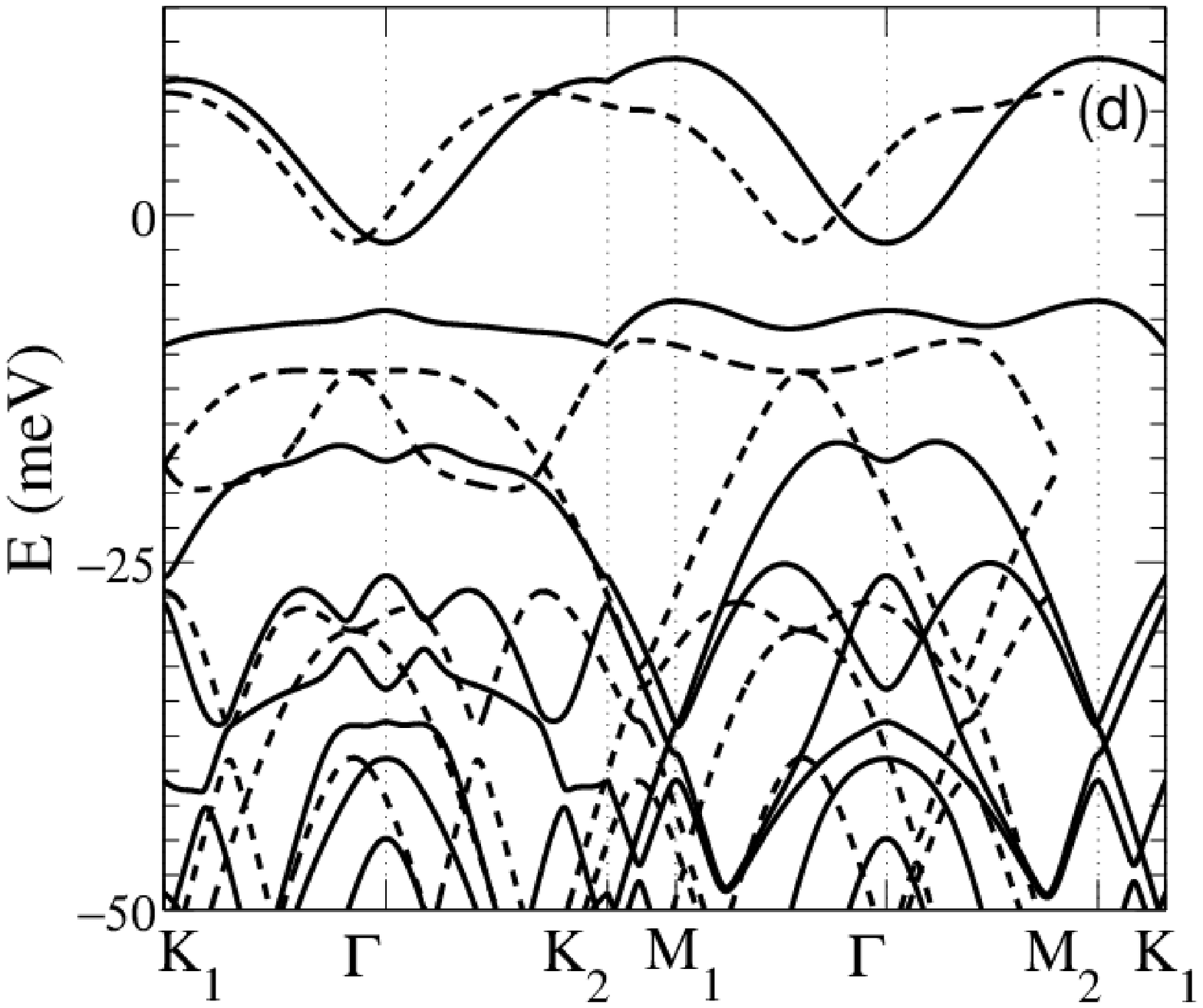}
\end{array}$
\end{center}
\caption{Valence energy band structure of twisted WSe$2$ homobilayer at $\theta=2^{\circ}$ under a uniaxial strain of $\epsilon_{xx}=-1\%$ (a), $\epsilon_{xx}=1\%$ (b) a shear strain $\epsilon_{xy}=-1\%$ (c) and $\epsilon_{xy}=1\%$ (d). The solid lines (dashed lines) correspond to the strained (unstrained) lattice.  Calculations are done for the relaxed lattice with $w=1.1$ meV, $w^{\prime}=0$, and $w^{\prime\prime}=9.7$ meV.}
\label{TMDfig2}
\end{figure}

The strain dependence of the effective velocity of t-BTMD (Eq. [\ref{v*TMD}]) shows that the lowest bands could be flatten by strain for a twist angle above $\theta_c$.
The strain-induced correction to the $v_x$ velocity component is
\begin{widetext}
\begin{eqnarray}
\Delta v^{\prime}_{x}&=&\xi\frac{v_F}{\langle\Psi|\Psi\rangle}\left\{\left(\frac {\hbar v_F k_{\theta}w}{X_0}\right)^2\frac 1{k_{\theta} a}
\left[-8\pi \epsilon_{xy}-Y\left(8\pi-6\sqrt{3}\beta\right)\epsilon_{xy}\right]\right.\nonumber\\
 &+&2\left. a_+\frac{\hbar v_F}a \left(\frac w{X_0}\right)^2\left[
 4\pi\left(1+Y\right)\epsilon_{xx}-4\pi Y\epsilon_{yy}-3\sqrt{3} \beta \left(\epsilon_{xx}-\epsilon_{yy}\right)\right]\right\},
\end{eqnarray}
\end{widetext}
where $X_0=\gamma-\left(\hbar v_F k_{\theta}\right)^2$, $Y=\frac{\left(\hbar v_F k_{\theta}\right)^2}{X_0}$, $a_+=U_++m/2$, and $U_+=\left(U_c+U_v\right)/2$.\newline
In the case of WSe$_2$, we take  for the rigid lattice, $w=w^{\prime}=w^{\prime\prime}=9.7$ meV, $\frac{\hbar v_F}a\sim 1.1$ eV \cite{Bi}, $\gamma\sim -1.1 10^{-3}$ $(eV)^2$, which could be neglected in the expression of $X_0$ and $Y$.\

Under a shear strain  $\Delta v^{\prime}_{x}$ reduces to
\begin{eqnarray}
\Delta v^{\prime}_{x}=-\frac{v_F}{\langle\Psi|\Psi\rangle}\xi\frac{9\sqrt{3}}{2\pi \theta}\beta\left(\frac w{\hbar v_F k_{\theta}}\right)^2
 \epsilon_{xy},
\end{eqnarray}
which corresponds, for $\theta=3^{\circ}$ to a correction of 
$\Delta v^{\prime}_{x}\sim - 0.2\,\epsilon_{xy} v_F$.\newline

Taking a strain component $\epsilon_{xx}$ and $\theta=3^{\circ}$, yields to 
\begin{eqnarray}
\Delta v^{\prime}_{x}=\frac{v_F}{\langle\Psi|\Psi\rangle}6\sqrt{3}\beta a_+ \frac{\hbar v_F}a \left(\frac w{X_0}\right)^2\epsilon_{xx}\sim 0.54 \,\epsilon_{xx} v_F\nonumber\\
\end{eqnarray}
The width of the lowest-energy flat bands of t-BTMD is, then, not strongly affected by the strain as depicted in Fig. \ref{TMDfig2}.

%\begin{widetext}

%%%%%%%%%% Merge with supplemental materials %%%%%%%%%%

\end{document}